\documentclass[%
 reprint,
 amsmath,amssymb,
 aps, nofootinbib
]{revtex4-1}
\usepackage{graphicx}
\usepackage{dcolumn}
\usepackage{bm}

\usepackage{amsmath,amssymb}
\usepackage{slashed}
\usepackage{graphicx}
\usepackage{dsfont}
\usepackage{bm}
\usepackage[top=1.0in,bottom=1.0in,left=1.0in,right=1.0in]{geometry}
\usepackage[colorlinks,linkcolor=blue,citecolor=blue]{hyperref}
\usepackage{CJKutf8}
\usepackage[caption=false]{subfig}

\begin{document}
\setlength{\oddsidemargin}{0.5cm}
\setlength{\topmargin}{-0.1cm}
\setlength{\textheight}{21cm}
\setlength{\textwidth}{15cm}
\newcommand{\be}{\begin{equation}}
\newcommand{\ee}{\end{equation}}
\newcommand{\bea}{\begin{eqnarray}}
\newcommand{\eea}{\end{eqnarray}}
\newcommand{\ba}{\begin{eqnarray}}
\newcommand{\ea}{\end{eqnarray}}

\newcommand{\fslash}{\hspace{-1.4ex}/\hspace{0.6ex} }
\newcommand{\Dslash}{D\hspace{-1.6ex}/\hspace{0.6ex} }
\newcommand{\Wslash}{W\hspace{-1.6ex}/\hspace{0.6ex} }
\newcommand{\pslash}{p\hspace{-1.ex}/\hspace{0.6ex} }
\newcommand{\kslash}{k\hspace{-1.ex}/\hspace{0.6ex} }
\newcommand{\underkslash}{{\underline k}\hspace{-1.ex}/\hspace{0.6ex} }
\newcommand{\epslash}{{\epsilon\hspace{-1.ex}/\hspace{0.6ex}}}
\newcommand{\partslash}{\partial\hspace{-1.6ex}/\hspace{0.6ex} }

\newcommand{\nn}{\nonumber}
\newcommand{\Tr}{\mbox{Tr}\;}
\newcommand{\tr}{\mbox{tr}\;}
\newcommand{\ket}[1]{\left|#1\right\rangle}
\newcommand{\bra}[1]{\left\langle#1\right|}
\newcommand{\rhoraket}[3]{\langle#1|#2|#3\rangle}
\newcommand{\brkt}[2]{\langle#1|#2\rangle}
\newcommand{\pdif}[2]{\frac{\partial #1}{\partial #2}}
\newcommand{\pndif}[3]{\frac{\partial^#1 #2}{\partial #3^#1}}
\newcommand{\pbm}[1]{\protect{\bm{#1}}}
\newcommand{\avg}[1]{\left\langle #1\right\rangle}
\newcommand{\vnabla}{\mathbf{\nabla}}
\newcommand{\notes}[1]{\fbox{\parbox{\columnwidth}{#1}}}
\newcommand{\pair}{\raisebox{-7pt}{\includegraphics[height=20pt]{pair0.pdf}}}
\newcommand{\paircrs}{\raisebox{-7pt}{\includegraphics[height=20pt]{pair0cross.pdf}}}
\newcommand{\paircc}{\raisebox{-7pt}{\includegraphics[height=20pt]{pair0cc.pdf}}}
\newcommand{\paircrscc}{\raisebox{-7pt}{\includegraphics[height=20pt]{pair0crosscc.pdf}}}
\newcommand{\pairloop}{\raisebox{-7pt}{\includegraphics[height=20pt]{pairloop.pdf}}}
\newcommand{\pairloopf}{\raisebox{-7pt}{\includegraphics[height=20pt]{pairloop4.pdf}}}
\newcommand{\pairlooph}{\raisebox{-7pt}{\includegraphics[height=20pt]{pair2looph.pdf}}}



\title{Tomography of the Rho meson  in the QCD Instanton Vacuum:\\ Transverse Momentum Dependent Parton Distribution Functions}

\author{Wei-Yang Liu}
\email{wei-yang.liu@stonybrook.edu}
\affiliation{Center for Nuclear Theory, Department of Physics and Astronomy, Stony Brook University, Stony Brook, New York 11794-3800, USA}

\author{Ismail Zahed }
\email{ismail.zahed@stonybrook.edu}
\affiliation{Center for Nuclear Theory, Department of Physics and Astronomy, Stony Brook University, Stony Brook, New York 11794-3800, USA}

\date{\today}
\begin{abstract}
We analyse the rho meson unpolarized and polarized transverse momentum dependent parton distribution functions (TMDPDFs) in the instanton liquid model (ILM). The corresponding TMDs in ILM are approximated by a constituent quark beam function in the leading Fock state multiplied by a rapidity-dependent factor resulting from the staple-shaped Wilson lines, for fixed longitudinal momentum, transverse separation, and rapidity. At the resolution of the ILM, all of the rho meson TMDs
are symmetric in parton $x$ for fixed transverse momentum, and Gaussian-like in the transverse momentum for fixed parton $x$. The latter is a direct consequence of the profiling of the  quark zero modes in the ILM. The evolved TMDs at higher rapidity using the Collins-Soper kernel, and higher resolution using the renormalization group (RG), show substantial skewness towards low parton $x$.
\end{abstract}

\maketitle

\section{Introduction}
Fast moving hadrons are composed of quarks and gluons (partons). The partonic structure of hadrons is probed through inclusive and semi-inclusive deep inelastic scattering experiments~\cite{Aschenauer:2019kzf}. Transverse momentum distributions (TMDs) for a given hadron, map the probability distribution of
a parton in both transverse momentum 
and longitudinal momentum fraction $x$~\cite{Boussarie:2023izj}. They reduce to the uni-modular parton distribution functions (PDFs) after transverse integration~\cite{Blumlein:2012bf}.

There is continous effort both experimentally~\cite{Bacchetta:2019sam,Vladimirov:2019bfa,Scimemi:2017etj,Cerutti:2022lmb,Bacchetta:2017gcc,Bacchetta:2022awv,Moos:2023yfa} and theoretically including instanton-based nonperturbative model \cite{Liu:2025mbl,Liu:2024sqj} and large scale lattice simulations~\cite{LatticeParton:2023xdl,LatticePartonLPC:2022eev,Deng:2022gzi,Ebert:2018gzl,Avkhadiev:2023poz,Schlemmer:2021aij}, to quantify these non-perturbative functions. Their understanding is part of a large experimental effort at COMPASS at CERN, JLAB in
Virginia, LHC at CERN and the future EIC at BNL. Currently, TMDPDFs are empirically extracted from 
Drell-Yan (DY) processes and semi-inclusive deep inelastic scattering (SIDIS), with small final hadron transverse momentum~\cite{Rogers:2015sqa}.

In the leading twist approximation, TMDs are defined as correlation functions of bilocal quarks with stapled light-like Wilson lines that are lined up 
on 
the light cone.  Recent lattice developments using the large momentum effective theory (LaMET) have proven useful for their possible Euclidean extraction and matching using quasi-TMDs, where the staple-shaped Wilson line (soft function) can be related to a mesonic form factor~\cite{LatticeParton:2020uhz}.

Spin-0 targets are characterized by 
one T-even TMD (unpolarized function) and one T-odd (Boer-Mulders function). Specifically, the T-even one reduces to the unpolarized parton distribution. 
Spin-$\frac 12$ targets are characterized 
by six T-even TMDs and two T-odd TMDs~\cite{Boussarie:2023izj,Pasquini:2019evu}. The T-odd TMDs are known as the Sivers and Boer-Mulders functions, while the T-even TMDs are tied to the standard unpolarized, helicity and transversity PDFs. Spin-1 targets do not have Sivers and Boer-Mulders functions. Alternatively, they have three additional T-even TMDs related to unpolarized partons in a tensor-polarized target, that amount to an additional PDF upon transverse integration~\cite{Ninomiya:2017ggn}.  The purpose of this paper is to evaluate the rho meson unpolarized and polarized  TMDs  at low resolution, using the instanton liquid model (ILM) description of the QCD vacuum. It is a follow up on our recent study of the spin-0 TMDs for the pion and kaon in the ILM~\cite{Liu:2025mbl}.


The Yang-Mills vacuum at high resolution is dominated by UV gauge fluctuations.
When the gradient flow cooling is applied to these configurations~\cite{Michael:1994uu,Michael:1995br,Leinweber:1999cw,Bakas:2010by,Biddle:2018bst,Hasenfratz:2019hpg,Athenodorou:2018jwu,Biddle:2020eec}, a smoother landscape in the IR emerges. 
This landscape is dominated by 
topologically active configurations known as pseudoparticles (instantons and and anti-instantons), with
a mean size $\rho \sim \frac 13\, \rm fm$, and a mean density (tunneling rate)
$n_{I+A}\sim 1/{\rm fm}^{4}$. These features are well captured by the ILM~\cite{ Diakonov:1985eg,Liu:2023fpj,Liu:2023yuj,Shuryak:2022wtk,Shuryak:2022thi,Shuryak:2021hng,Shuryak:2021fsu,Liu:2024rdm,Liu:2024jno,Zahed:2022wae,Zahed:2021fxk}. For more comprehensive review, see also \cite{Diakonov:2002fq,Schafer:1996wv,Liu:2025ldh} (and references therein).

The organization of the paper is as follow: In section~\ref{SECII} we outline  the general aspects of the rho TMDs,  and develop their analysis in the ILM,
at low resolution. We also derive their relationships to the rho PDFs and 
sum rules. In section~\ref{SECIII} we show how to implement the soft subtraction
scheme on the TMD distributions, and carry their double evolution in scale and rapidity. Our conclusions are in section~\ref{SECIV}. A number of Appendices are added  to detail and support the  arguments in the main text.

\section{Rho TMDs}
\label{SECII}

\subsection{General}
The leading twist-2 TMDs describe the unpolarized, helicity and several
transversity parton distributions in the transverse plane
\begin{widetext}
\begin{align}
\label{tmd}
    q_\rho(x,k_\perp)=&\int_{-\infty}^\infty\frac{dz^-}{2\pi}\int\frac{d^2b_\perp}{(2\pi)^2}e^{ix P^+z^--ik_\perp\cdot b_\perp}\langle \rho|\bar{\psi}(0) \gamma^+W^{(\pm)}[0,0_\perp; z^-,b_\perp]\psi(z^-,b_\perp)|\rho\rangle \nonumber\\
    \Delta q_\rho(x,k_\perp)=&\int_{-\infty}^\infty\frac{dz^-}{2\pi}\int\frac{d^2b_\perp}{(2\pi)^2}e^{ix P^+z^--ik_\perp\cdot b_\perp}\langle \rho|\bar{\psi}(0) \gamma^+\gamma^5W^{(\pm)}[0,0_\perp; z^-,b_\perp]\psi(z^-,b_\perp)|\rho\rangle \nonumber\\
    \delta q_\rho^\alpha(x,k_\perp)=&\int_{-\infty}^\infty\frac{dz^-}{2\pi}\int\frac{d^2b_\perp}{(2\pi)^2}e^{ix P^+z^--ik_\perp\cdot b_\perp}\langle \rho|\bar{\psi}(0)i\sigma^{\alpha+}\gamma^5 W^{(\pm)}[0,0_\perp; z^-,b_\perp]\psi(z^-,b_\perp)|\rho\rangle
\end{align}
\end{widetext}
Here the hadronic state label includes the quantum numbers carried by the
on-shell vector meson of mass $m_\rho$., i.e.  $\rho(p,\lambda)$. The rho kinematics is represented by the longitudinal light front unit vectors $n=(1,0^-,0_\perp)$ and $\bar n=(0^+,1,0_\perp)$ in light-cone signature, and the transverse component. The hadronic momentum with energy $E_{\vec{p}}=\sqrt{m_\rho^2+\vec{p}^2}$ and $3$-momentum $\vec{p}$,  can be decomposed 
\begin{equation}
\begin{aligned}
    p^\mu=&p^+\bar n^\mu-\frac {m^2_\rho}{2p^+} n^\mu+p^\mu_\perp
\end{aligned}
\end{equation}

The Wilson lines in the matrix elements  \eqref{tmd} enforce gauge-invariance.
They are  different in semi-inclusive deep inelastic scattering (SIDIS) process (space-like), and in Drell-Yan process (time-like) \cite{Kumano:2020ijt}.
The resummation in  the SIDIS processes of the intermediate gluons, yields the Wilson line defined as $W^{(+)}$. The resummations in the Drell-Yan processes of the intermediate gluons, leads the Wilson line defined as $W^{(-)}$ \cite{Angeles-Martinez:2015sea,GrossePerdekamp:2015xdx,Aidala:2012mv,Barone:2010zz,DAlesio:2007bjf},
\begin{widetext}
\begin{equation}
\label{TMD_wilson}
    W^{(\pm)}[0,0_\perp; z^-,b_\perp] = W_n[0,0_\perp ; \pm\infty,0_\perp ] W_\perp[ \pm\infty,0_\perp ; 
    \pm\infty, b_\perp] W_n[ \pm\infty,b_\perp ; z^-,b_\perp].
\end{equation}
\end{widetext}
with the generic gauge link defined as
\begin{equation}
    W[x;y]=\mathcal{P}\exp\left[ig\int_{x}^{y} dz_\mu A^\mu(z)\right]
\end{equation}

\begin{figure*}
\centering
%
%
\subfloat[\label{fig:rhof1_contour}]{\includegraphics[width=0.35\linewidth]{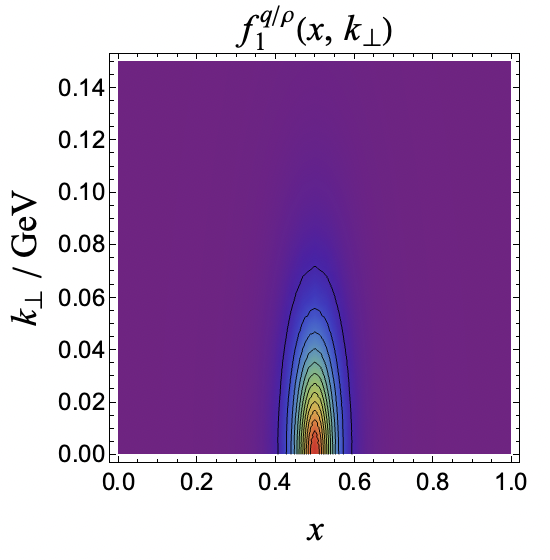}}
\hfill
\subfloat[\label{fig:rhof1LT_contour}]{\includegraphics[width=0.35\linewidth]{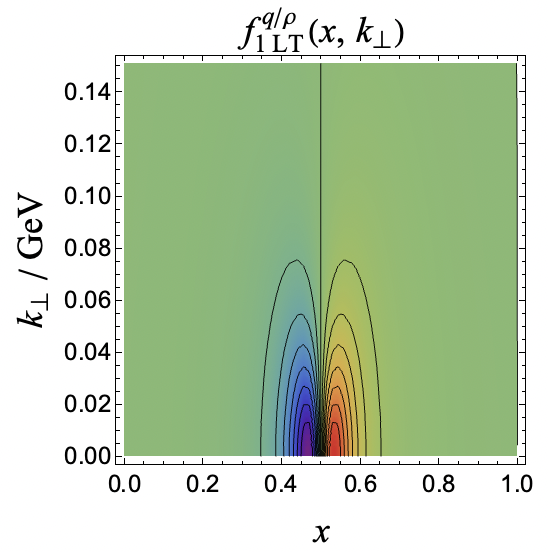}}
\hfill
\subfloat[\label{fig:rhof1_3d}]{\includegraphics[width=0.35\linewidth]{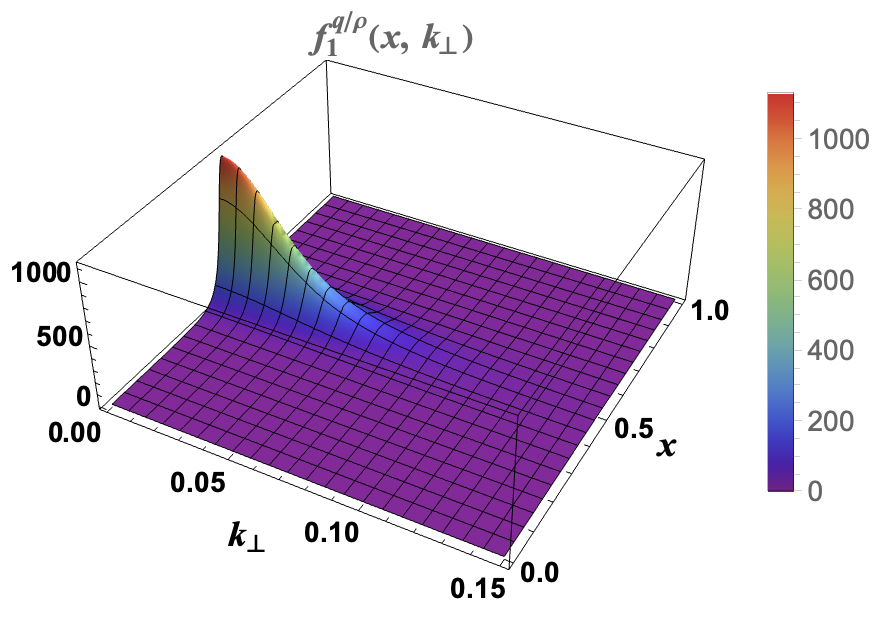}}
\hfill
\subfloat[\label{fig:rhof1LT_3d}]{\includegraphics[width=0.35\linewidth]{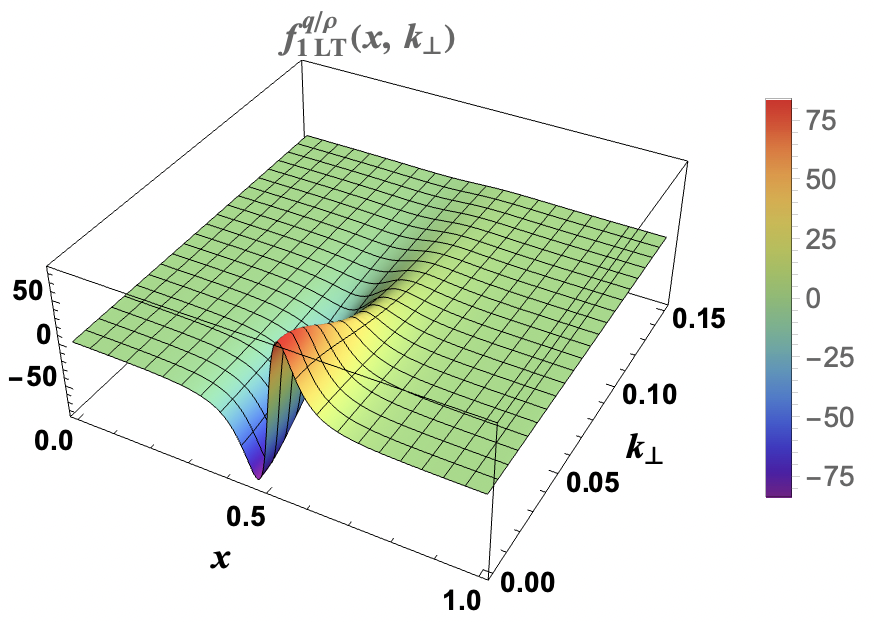}}
\hfill
\subfloat[\label{fig:rhof1_x}]{\includegraphics[width=0.35\linewidth]{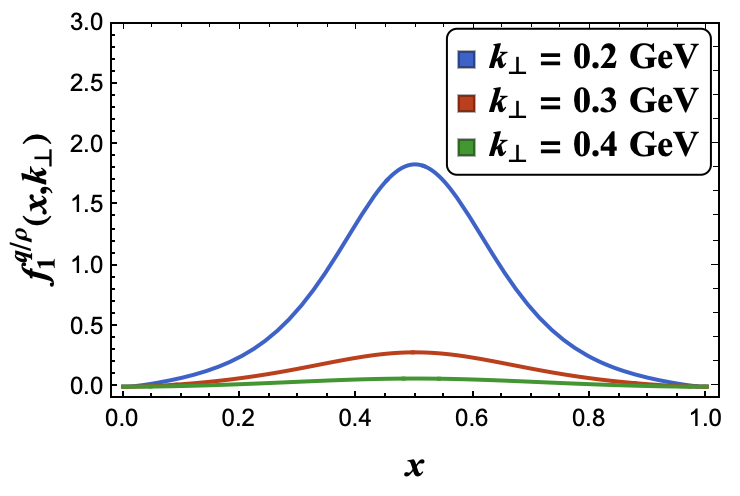}}
\hfill
\subfloat[\label{fig:rhof1LT_x}]{\includegraphics[width=0.35\linewidth]{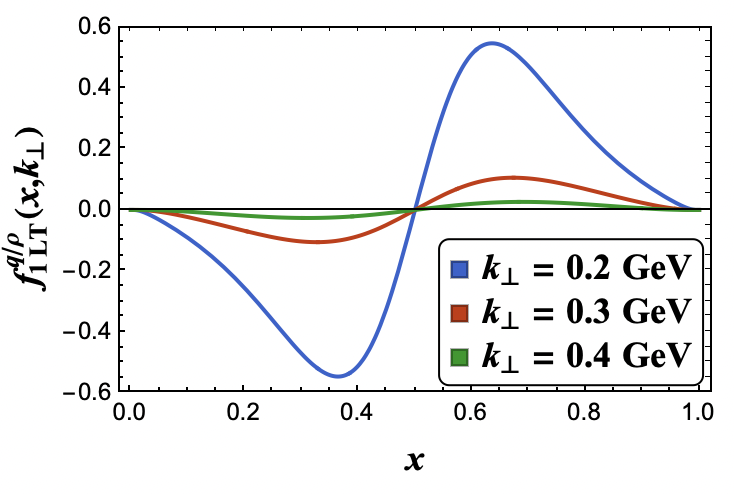}}
\hfill
\subfloat[\label{fig:rhof1_k}]{\includegraphics[width=0.35\linewidth]{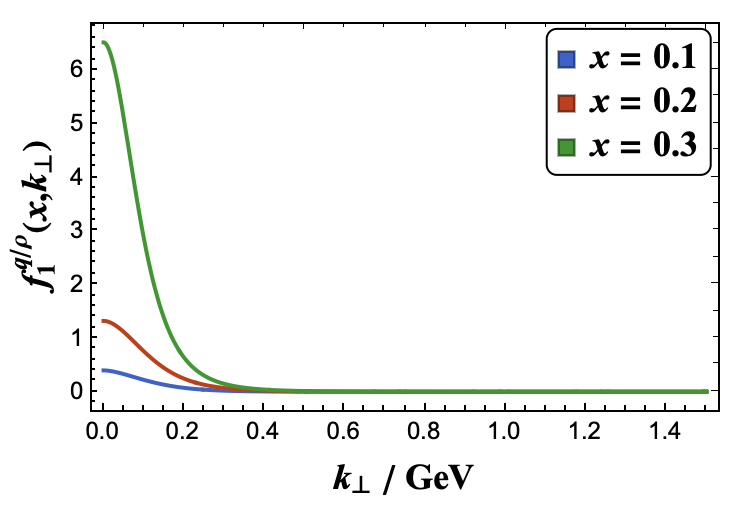}}
\hfill
\subfloat[\label{fig:rhof1LT_k}]{\includegraphics[width=0.35\linewidth]{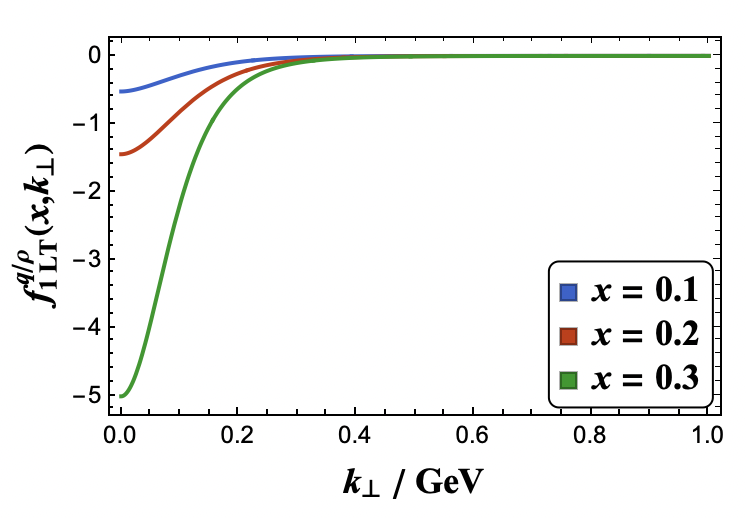}}
\caption{Rho meson TMD $f_1^{q/\rho}$ (left) \eqref{F1QR}  and $f_{1\mathrm{LT}}^{q/\rho}$ (right) 
\eqref{F1LTQR} beam functions with only constituent quark contribution at low resolution $\mu=1/\rho$
with $\rho=0.313$ fm: (a,b) are the density plots, (c,d) the 3D plots,
(e,f) the longitudinal momentum dependence for fixed $k_\perp$, and (g,h) the transverse momentum dependent plots for fixed $x$. The rho LFWF parameters are $C_{\rho}=1.9888$, $m_\rho=791.0$ MeV, $M=398.17$ MeV.}
\label{fig:rho_TMDf1}
\end{figure*}

\begin{figure*}
\centering
%
%
\subfloat[\label{fig:rhof1LL_contour}]{\includegraphics[width=0.35\linewidth]{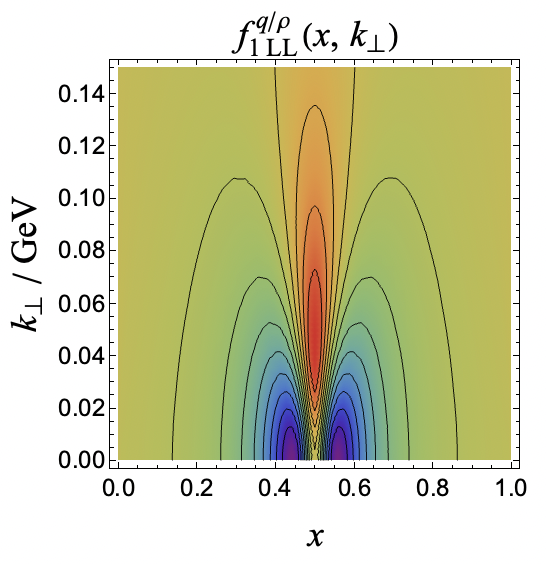}}
\hfill
\subfloat[\label{fig:rhof1TT_contour}]{\includegraphics[width=0.35\linewidth]{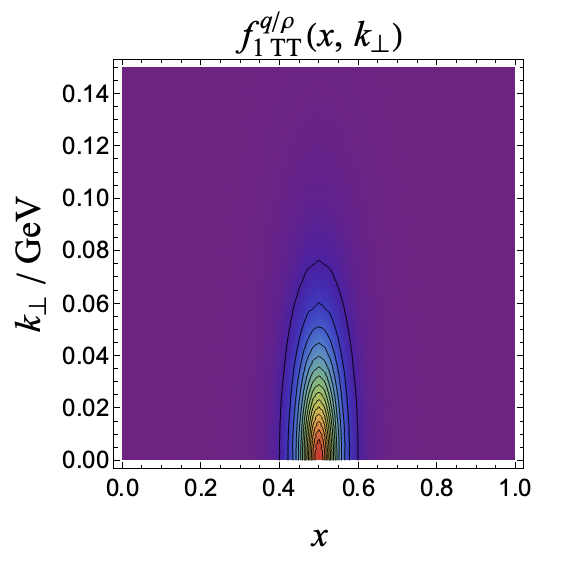}}
\hfill
\subfloat[\label{fig:rhof1LL_3d}]{\includegraphics[width=0.35\linewidth]{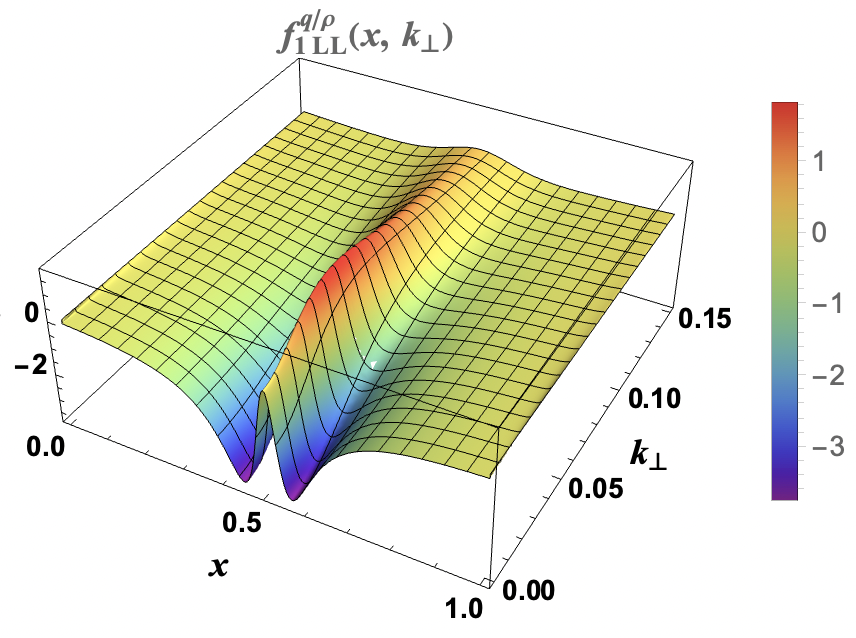}}
\hfill
\subfloat[\label{fig:rhof1TT_3d}]{\includegraphics[width=0.35\linewidth]{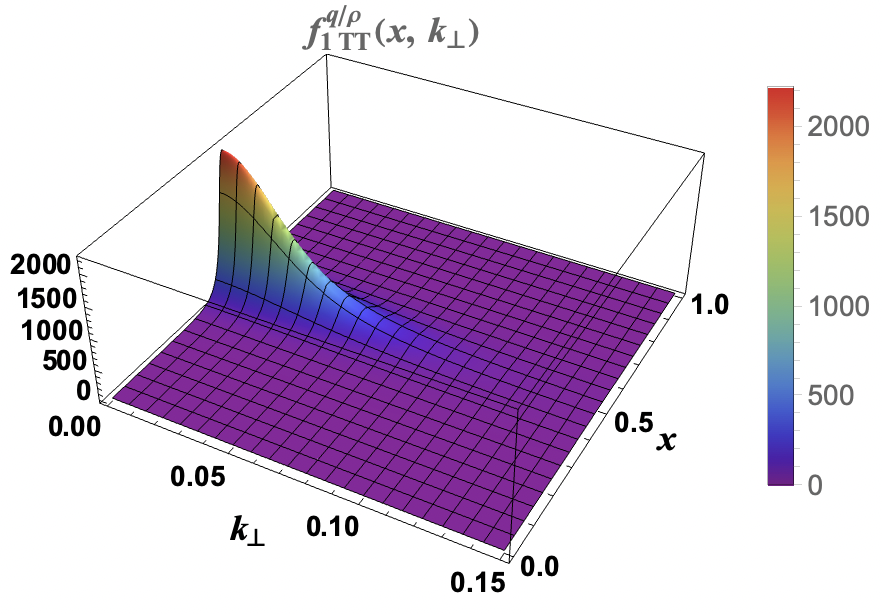}}
\hfill
\subfloat[\label{fig:rhof1LL_x}]{\includegraphics[width=0.35\linewidth]{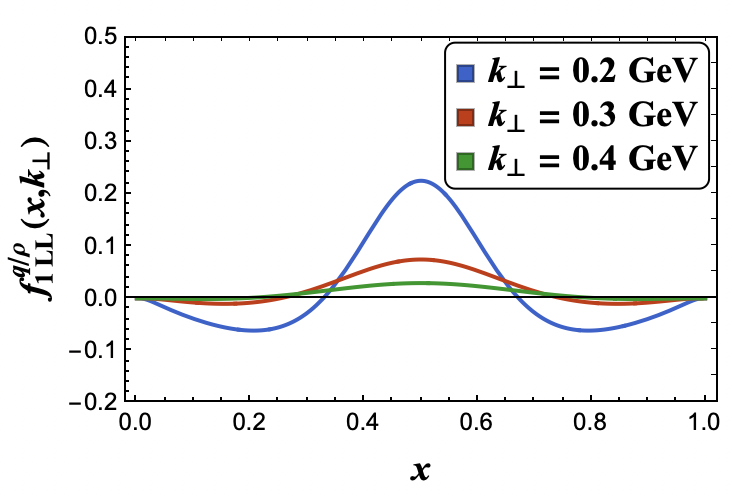}}
\hfill
\subfloat[\label{fig:rhof1TT_x}]{\includegraphics[width=0.35\linewidth]{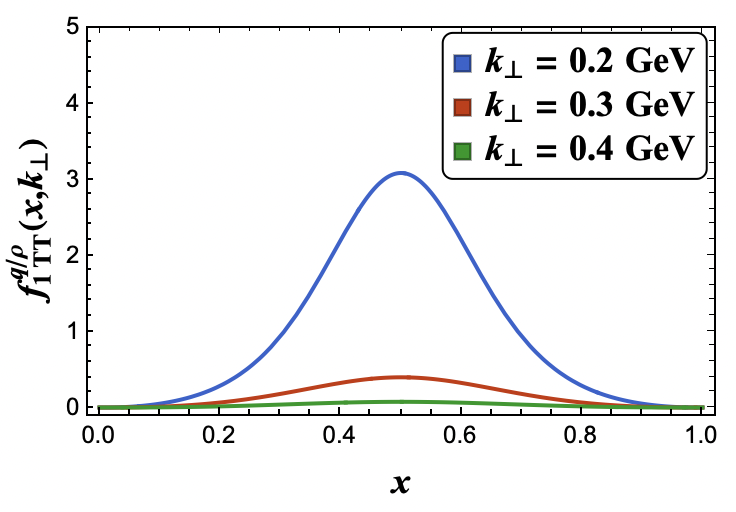}}
\hfill
\subfloat[\label{fig:rhof1LL_k}]{\includegraphics[width=0.35\linewidth]{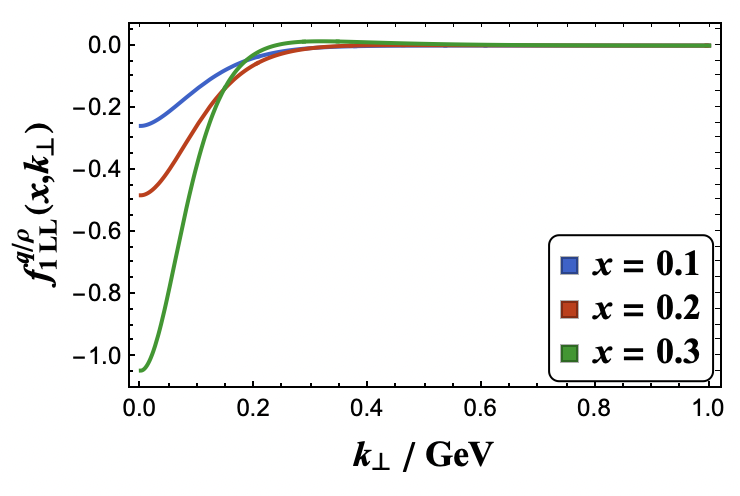}}
\hfill
\subfloat[\label{fig:rhof1TT_k}]{\includegraphics[width=0.35\linewidth]{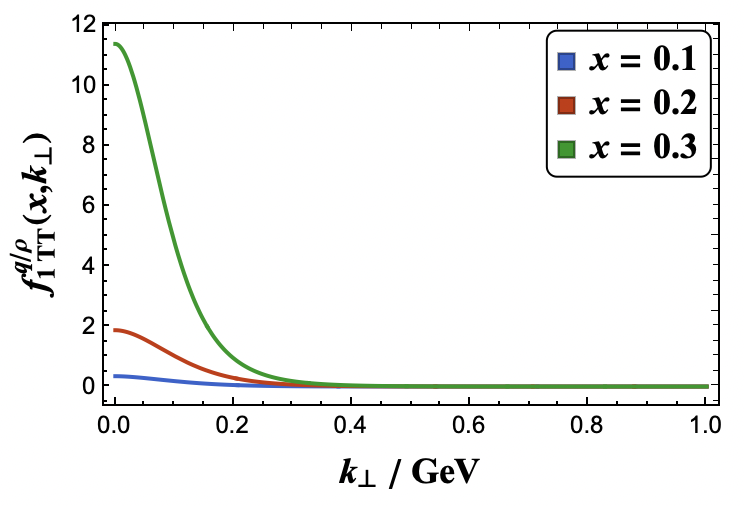}}
\caption{Rho meson tensor-polarized TMD $f_{1LL}^{q/\rho}$ (left) \eqref{F1LLQR}  and $f_{1TT}^{q/\rho}$ (right) 
\eqref{F1TTQR} beam functions with only constituent quark contribution at low resolution $\mu=1/\rho$
with $\rho=0.313$ fm: (a,b) are the density plots, (c,d) the 3D plots,
(e,f) the longitudinal momentum dependence for fixed $k_\perp$, and (g,h) the transverse momentum dependent plots for fixed $x$. The rho LFWF parameters are $C_{\rho}=1.9888$, $m_\rho=791.0$ MeV, $M=398.17$ MeV.}
\label{fig:rho_TMDf1TT}
\end{figure*}

\begin{figure*}
\centering
%
%
\subfloat[\label{fig:rhog1_contour}]{\includegraphics[width=0.35\linewidth]{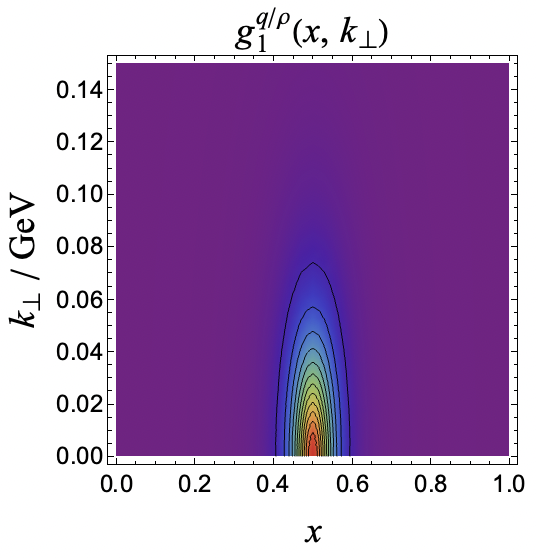}}
\hfill
\subfloat[\label{fig:rhog1T_contour}]{\includegraphics[width=0.35\linewidth]{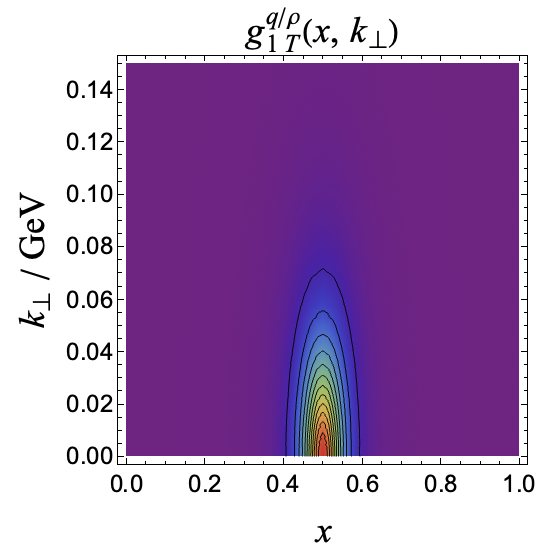}}
\hfill
\subfloat[\label{fig:rhog1_3d}]{\includegraphics[width=0.35\linewidth]{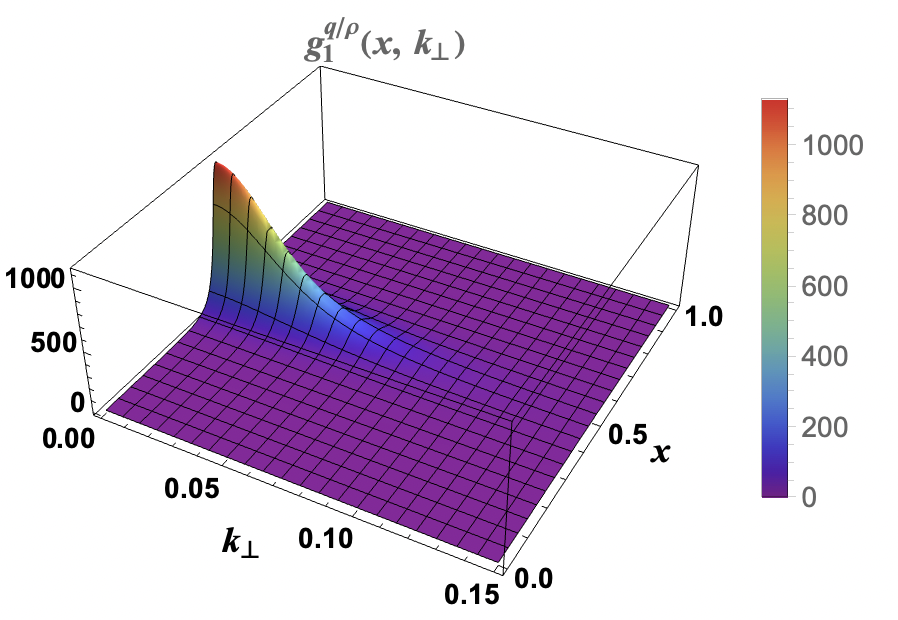}}
\hfill
\subfloat[\label{fig:rhog1T_3d}]{\includegraphics[width=0.35\linewidth]{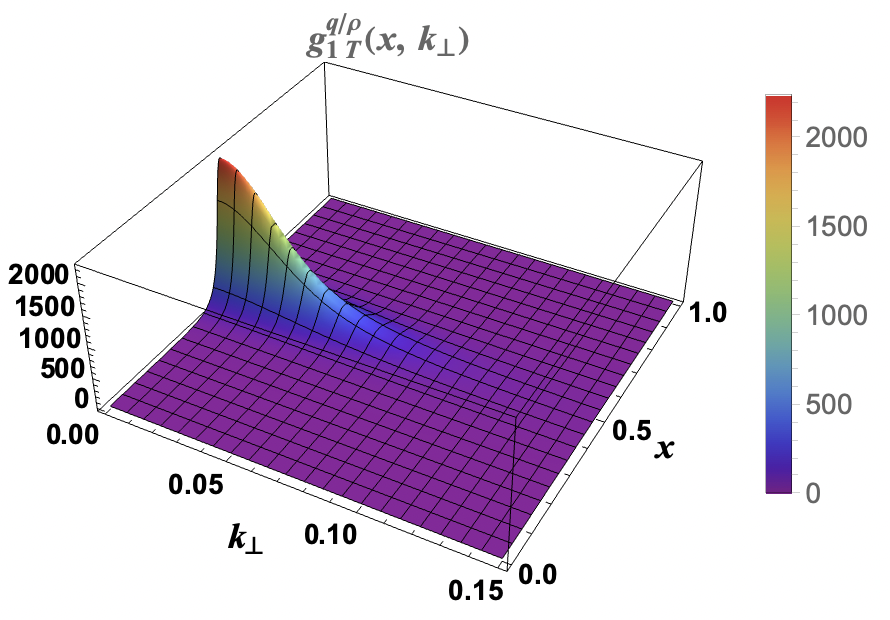}}
\hfill
\subfloat[\label{fig:rhog1_x}]{\includegraphics[width=0.35\linewidth]{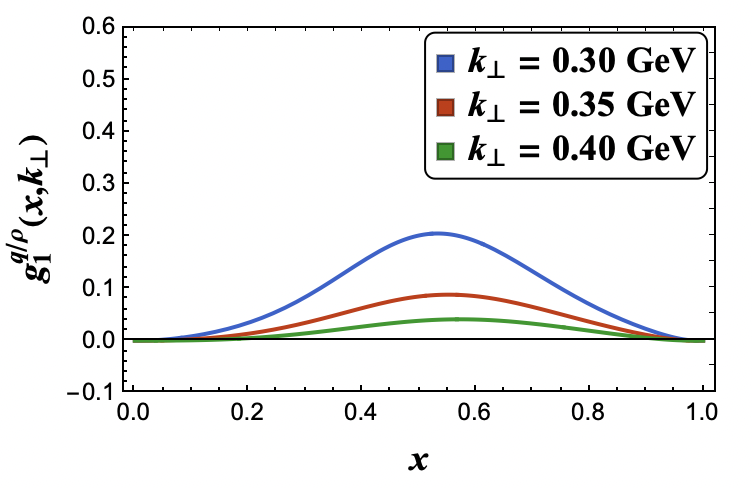}}
\hfill
\subfloat[\label{fig:rhog1T_x}]{\includegraphics[width=0.35\linewidth]{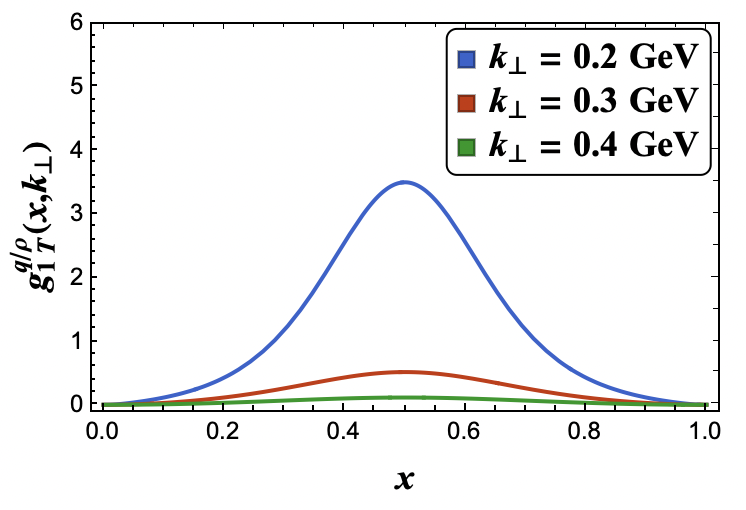}}
\hfill
\subfloat[\label{fig:rhog1_k}]{\includegraphics[width=0.35\linewidth]{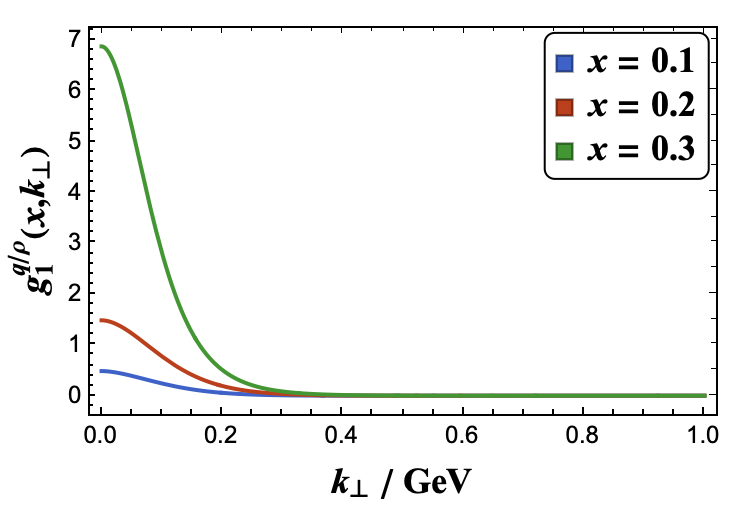}}
\hfill
\subfloat[\label{fig:rhog1T_k}]{\includegraphics[width=0.35\linewidth]{rhog1_TMD_k.png}}
\caption{Rho meson TMD $g_1^{q/\rho}$ (left) \eqref{G1QR} and $g_{1T}^{q/\rho}$ (right) 
\eqref{G1TQR} beam functions with only constituent quark contribution at low resolution $\mu=1/\rho$
with $\rho=0.313$ fm: (a,b) are the density plots, (c,d) the 3D plots,
(e,f) the longitudinal momentum dependence for fixed $k_\perp$, and (g,h) the transverse momentum dependent plots for fixed $x$. The rho LFWF parameters are $C_{\rho}=1.9888$, $m_\rho=791.0$ MeV, $M=398.17$ MeV.}
\label{fig:rho_TMDg1}
\end{figure*}

\begin{figure*}
\centering
%
%
\subfloat[\label{fig:rhoh1_contour}]{\includegraphics[width=0.35\linewidth]{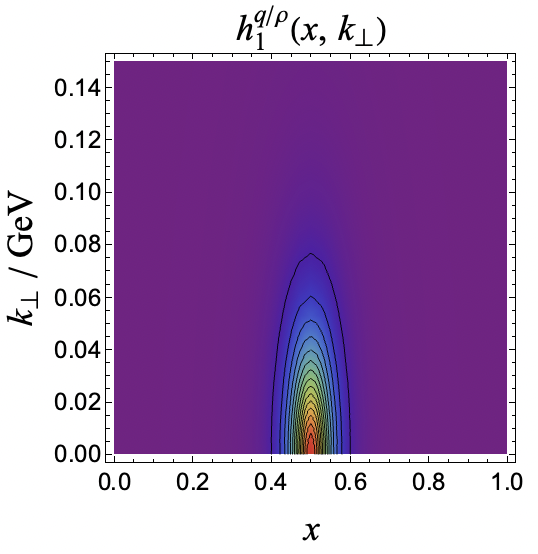}}
\hfill
\subfloat[\label{fig:rhoh1L_contour}]{\includegraphics[width=0.35\linewidth]{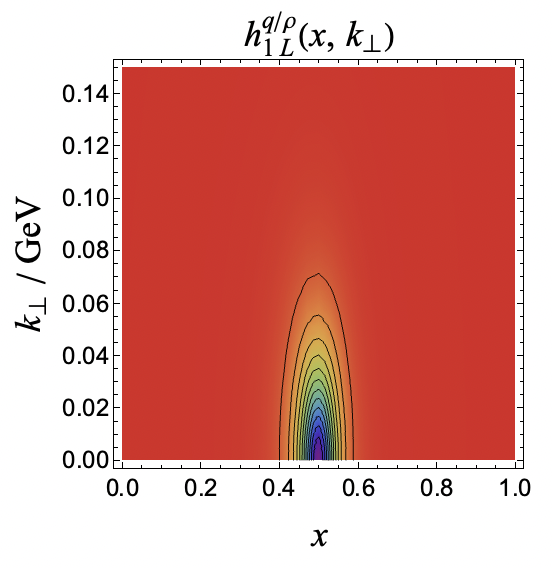}}
\hfill
\subfloat[\label{fig:rhoh1_3d}]{\includegraphics[width=0.35\linewidth]{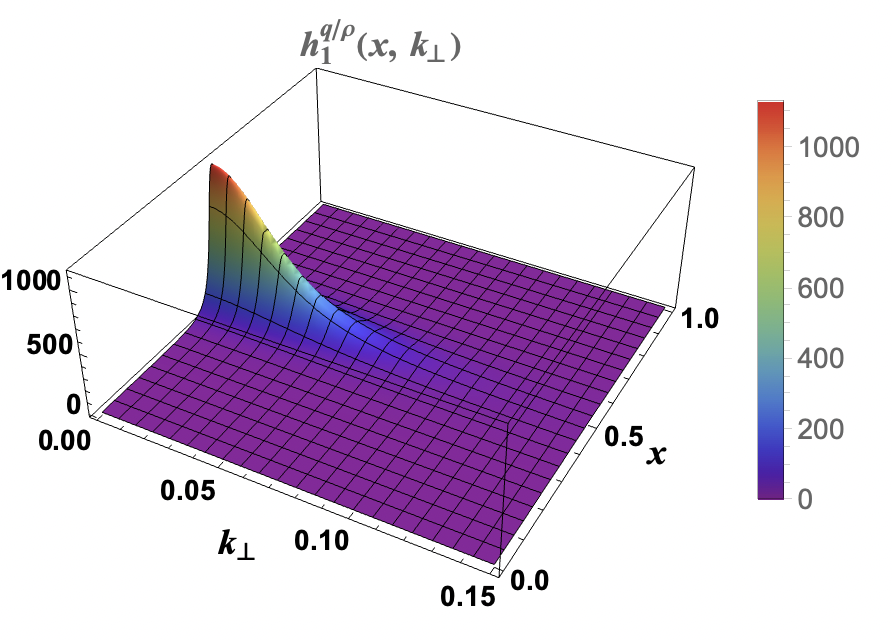}}
\hfill
\subfloat[\label{fig:rhoh1L_3d}]{\includegraphics[width=0.35\linewidth]{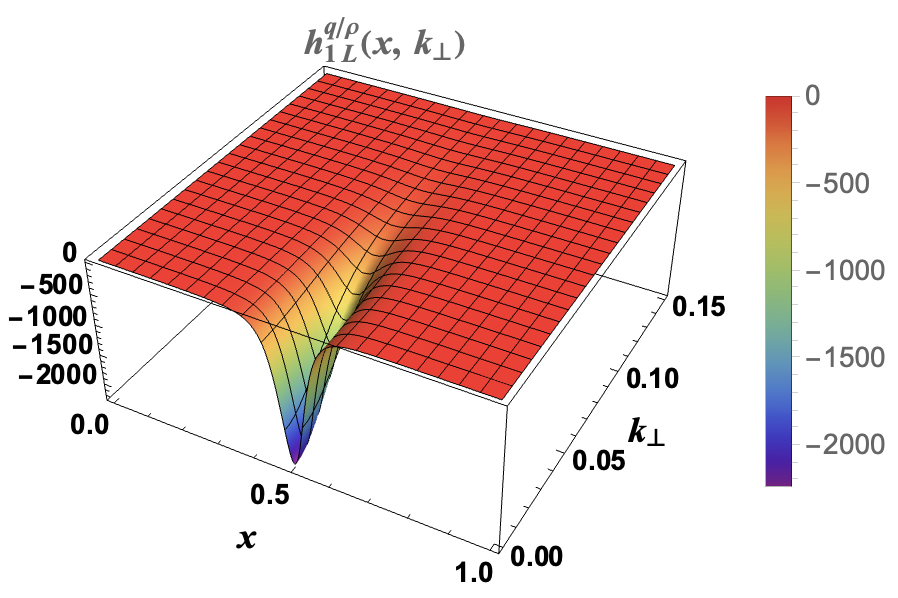}}
\hfill
\subfloat[\label{fig:rhoh1_x}]{\includegraphics[width=0.35\linewidth]{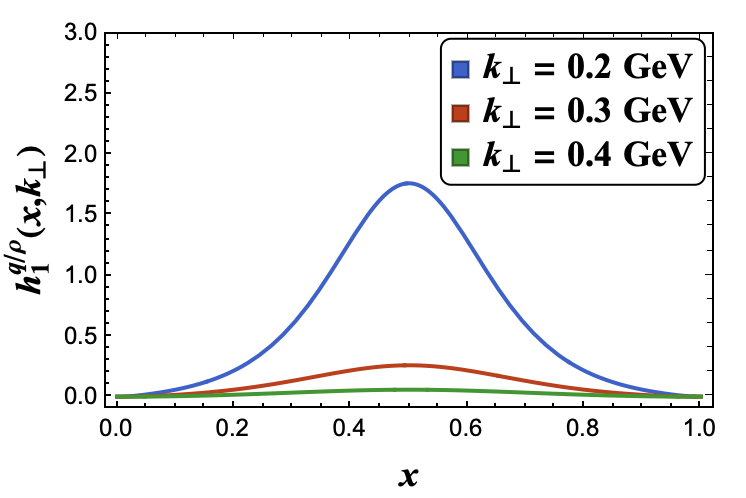}}
\hfill
\subfloat[\label{fig:rhoh1L_x}]{\includegraphics[width=0.35\linewidth]{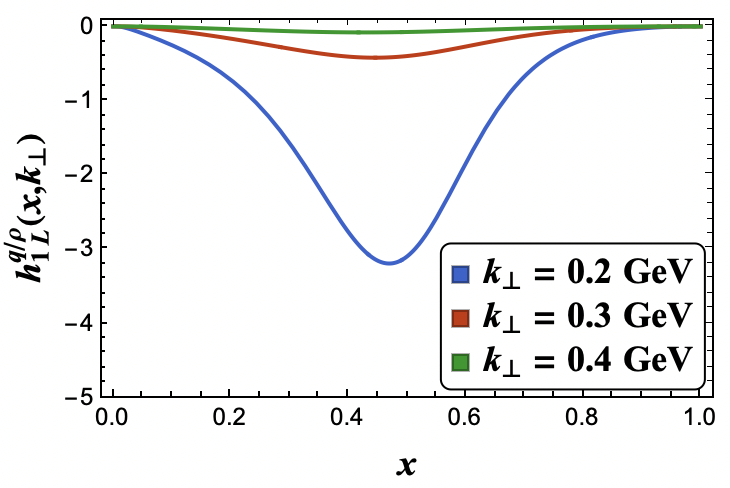}}
\hfill
\subfloat[\label{fig:rhoh1_k}]{\includegraphics[width=0.35\linewidth]{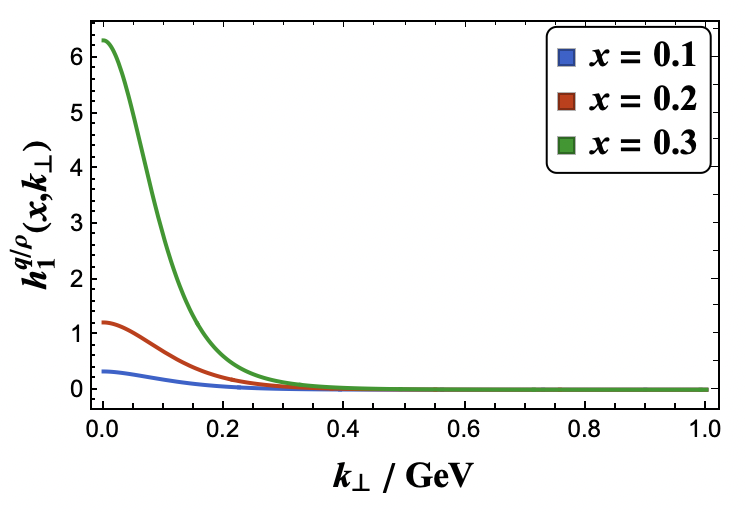}}
\hfill
\subfloat[\label{fig:rhoh1L_k}]{\includegraphics[width=0.35\linewidth]{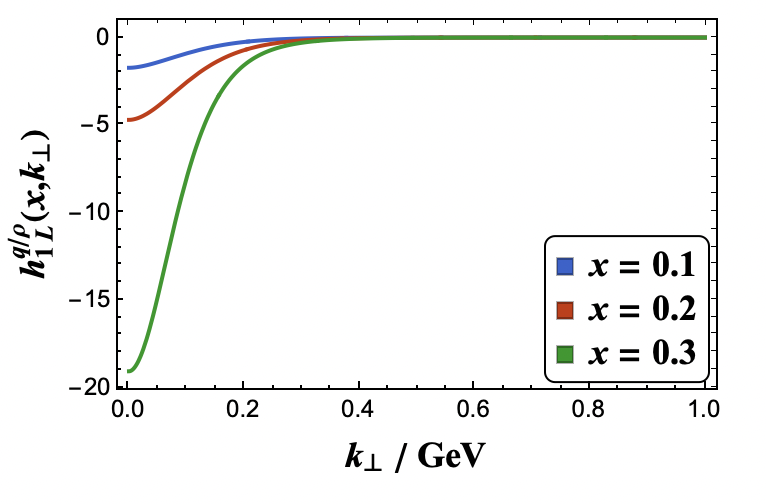}}
\caption{Rho meson TMD $h_1^{q/\rho}$ (left) \eqref{H1QR} and $h_{1L}^{q/\rho}$ (right) 
\eqref{H1LQR} beam functions with only constituent quark contribution at low resolution $\mu=1/\rho$
with $\rho=0.313$ fm: (a,b) are the density plots, (c,d) the 3D plots,
(e,f) the longitudinal momentum dependence for fixed $k_\perp$, and (g,h) the transverse momentum dependent plots for fixed $x$. The rho LFWF parameters are $C_{\rho}=1.9888$, $m_\rho=791.0$ MeV, $M=398.17$ MeV.}
\label{fig:rho_TMDh1}
\end{figure*}

\subsection{Momentum distribution}
The rho TMDs can be decomposed into several functions based on the quark spin direction compared to the polarization of the hadron states.  The notations follow the convention in PDFs. The functions $f$, $g$ and $h$ denote the quark polarization for being unpolarized ($\gamma^+$), longitudinally polarized ($\gamma^+\gamma^5$) and transversely polarized ($i\sigma^{\alpha+}\gamma^5$) respectfully. The lower index $1$ refers to the
leading twist, and the subscript $L$ and $T$ refer to the hadron polarization. 


For rho mesons, the  Lorentz decomposition makes use of  the spin quantization axis $\vec{s} = (s_\perp,s_L)$, and the
vector meson polarization $\lambda = 0, \pm1$ projected onto $\vec{s}$. With this in mind, the TMDs for vector mesons can be decomposed as~\cite{Shi:2022erw}
\begin{widetext}
\begin{align}
    q_\rho(x,k_\perp)=&~f^{q/\rho}_1(x,k_\perp)+S_{LL}f^{q/\rho}_{1LL}(x,k_\perp)+\frac{S_{LT}\cdot k_\perp}{m_\rho}f^{q/\rho}_{1LT}(x,k_\perp)+\frac{ S^{\alpha\beta}_{TT}k^\alpha_\perp k^\beta_\perp}{m^2_\rho}f^{q/\rho}_{1TT}(x,k_\perp) \\
    \Delta q_\rho(x,k_\perp)=&~\lambda \left[s_L g^{q/\rho}_{1}(x,k_\perp)+\frac{k_\perp\cdot s_\perp}{m_\rho}g^{q/\rho}_{1T}(x,k_\perp)\right] \\
    \delta q_\rho^\alpha(x,k_\perp)=&~\lambda\left[s^\alpha_\perp h^{q/\rho}_{1}(x,k_\perp)+s_L\frac{k^\alpha_\perp}{m_\rho}h^{q/\rho}_{1L}(x,k_\perp)+\frac{1}{2m_\rho^2}\left(s^\alpha_\perp k^2_\perp-2k_\perp\cdot s_\perp k^\alpha_\perp\right) h^{q/\rho}_{1T}(x,k_\perp)\right]
\end{align}
\end{widetext}
with the spin polarization tensors
\begin{align}
     S_{LL}=&(3\lambda^2-2)\left(\frac16-\frac12s^2_L\right) \\
     S^\alpha_{LT}=&(3\lambda^2-2)s_Ls_\perp^\alpha \\
     S^{\alpha\beta}_{TT}=&(3\lambda^2-2)\left(s_\perp^\alpha s_\perp^\beta-\frac12s^2_\perp\delta^{\alpha\beta}\right)
\end{align} 
There are three tensor polarized TMDs $f_{1LL}$, $f_{1LT}$, and $f_{1TT}$ that are specific to spin-one mesons.  As we noted earlier, for spin-one mesons there are nine leading-twist TMDs and four PDFs, if we restrict ourselves to the $T$-even distributions.

With the Fourier transform of a generic TMD $F^{q/\rho}(x,k_\perp)$ to the conjugated transverse coordinate space defined as,

\begin{equation}
    \tilde{F}^{q/\rho}(x,b_\perp)=\int\frac{d^2k_\perp}{(2\pi)^2}e^{ik_\perp\cdot b_\perp} F^{q/\rho}(x,k_\perp)
\end{equation}
the Lorentz decomposition of TMD distribution in $b_\perp$ space can be written as~\cite{Moos:2020wvd}
\begin{widetext}
\begin{align}
\label{tmd_b_2}
\tilde{q}_\rho(x,b_\perp)=&~\tilde{f}^{q/\rho}_1(x,b_\perp)+S_{LL}\tilde{f}^{q/\rho}_{1LL}(x,b_\perp)+iS_{LT}\cdot b_\perp m_\rho\tilde{f}^{q/\rho}_{1LT}(x,b_\perp)+ S^{\alpha\beta}_{TT}b^\alpha_\perp b^\beta_\perp m^2_\rho\tilde{f}^{q/\rho}_{1TT}(x,b_\perp) \\
    \Delta \tilde{q}_\rho(x,b_\perp)=&~\lambda \left[s_L \tilde{g}^{q/\rho}_{1}(x,b_\perp)+ib_\perp\cdot s_\perp m_\rho\tilde{g}^{q/\rho}_{1T}(x,b_\perp)\right] \\
    \delta \tilde{q}_\rho^\alpha(x,b_\perp)=&~\lambda\left[s^\alpha_\perp \tilde{h}^{q/\rho}_{1}(x,b_\perp)+is_Lb^\alpha_\perp m_\rho\tilde{h}^{q/\rho}_{1L}(x,b_\perp)-\frac{m_\rho^2}{2}\left(s^\alpha_\perp b^2_\perp-2b_\perp\cdot s_\perp b^\alpha_\perp\right) \tilde{h}^{q/\rho}_{1T}(x,b_\perp)\right]
\end{align}
\end{widetext}
The relations between the $b_\perp$ space TMD and the $k_\perp$ space TMDs,
are listed in Appendix \ref{sec:tmd_b}

\subsection{Rho TMD in ILM}
The computation of TMD quark distribution in the vector meson can be simply addressed by the light front wave functions (LFWFs). The softly factored contribution of the staple-shaped Wilson line $W^{(\pm)}$ at low resolution, will be addressed below. With this in mind, the rho meson TMD reads
\begin{widetext}
\begin{align}
\label{tmd_rho}
    &q_\rho(x,k_\perp)=\frac{1}{(2\pi)^3}\sum_{s_1,s_2}\Phi^\dagger_\rho(x,k_\perp,s_1,s_2)\Phi_\rho(x,k_\perp,s_1,s_2) \nonumber\\
    &\Delta q_\rho(x,k_\perp)=\frac{1}{(2\pi)^3}\sum_{s_1,s_1',s_2}\Phi^\dagger_\rho(x,k_\perp,s_1,s_2)\sigma^3_{s_1s_1'}\Phi_\rho(x,k_\perp,s_1',s_2) \nonumber\\
    &\delta q^{\alpha}_{\rho}(x,k_\perp)=\frac{1}{(2\pi)^3}\sum_{s_1,s_1',s_2}\Phi^\dagger_\rho(x,k_\perp,s_1,s_2)\sigma^\alpha_{s_1s_1'}\Phi_\rho(x,k_\perp,s_1',s_2)
\end{align}
With the vector meson wave functions polarized in the $\vec{s}$-direction.
Their general form follows from the derivation given in~\cite{Liu:2023fpj}, using an arbitrary spin quantization axis. The result for the longitudinal vector meson LFWF,  is
\begin{equation}
\begin{aligned}
\label{LLX}
    \Phi^0_{\rho}(x,k_\perp,s_1,s_2)
    &=\frac{1}{\sqrt{N_c}}\left[\frac{C_{\rho}}{\sqrt{x\bar{x}}\left(m^2_\rho-\frac{k^2_\perp+M^2}{x\bar{x}}\right)}\mathcal{F}\left(\frac{k_\perp}{\lambda_{\rho}\sqrt{x\bar x}}\right)\right]\varepsilon^\mu_{0}(p)\bar{u}_{s_1}(k_1)\gamma_{\mu}\tau^a v_{s_2}(k_2)
\end{aligned}
\end{equation}
For the transverse vector meson LFWFs, the results are
\begin{equation}
\begin{aligned}
\label{TTX}
    \Phi^\pm_{\rho}(x,k_\perp,s_1,s_2)
    &=\frac{1}{\sqrt{N_c}}\left[\frac{C_{\rho}}{\sqrt{x\bar{x}}\left(m^2_\rho-\frac{k^2_\perp+M^2}{x\bar{x}}\right)}\mathcal{F}\left(\frac{k_\perp}{\lambda_{\rho}\sqrt{x\bar x}}\right)\right]\varepsilon^\mu_{\pm}(p)\bar{u}_{s_1}(k_1)\gamma_{\mu}\tau^a v_{s_2}(k_2)
\end{aligned}
\end{equation}
Here $u_{s_1}(k_1)$ denotes the light front quark spinor with spin $s_1$ and  internal momentum $k^+_1=xp^+$, $k_{1\perp}=k_\perp$. Similarly, $v_{s_2}(k_2)$ denotes the light front anti-quark spinor with spin $s_2$ and internal momentum $k^+_2=\bar{x}p^+$, $k_{2\perp}=-k_\perp$, as detailed in Appendix \ref{Appx:LFspinor}. The polarization direction is encoded in the polarization vector $\varepsilon^\mu(p)$. The normalization condition for the light front wavefunctions, yields
\begin{equation}
\begin{aligned}
\label{LF_T}
C_{\rho_T}=&\left[\frac{1}{4\pi^2}\int_0^1dx\int_0^\infty dk^2_\perp\frac{k^2_\perp+M^2-2x\bar{x}k^2_\perp}{(x\bar{x}m_{\rho}^2-k_\perp^2-M^2)^2}\mathcal{F}^2\left(\frac{k_\perp}{\lambda_{\rho}\sqrt{x\bar x}}\right)\right]^{-1/2}\\
\end{aligned}
\end{equation}
and
\begin{equation}
\begin{aligned}
\label{LF_L}
C_{\rho_L}=&\left[\frac{1}{4\pi^2}\int_0^1dx\int_0^\infty dk^2_\perp\left(\frac1{m_\rho^2}+\frac{4x\bar{x}(k_\perp^2+M^2)}{(x\bar{x}m_{\rho}^2-k_\perp^2-M^2)^2}\right)\mathcal{F}^2\left(\frac{k_\perp}{\lambda_{\rho}\sqrt{x\bar x}}\right)\right]^{-1/2}\\
\end{aligned}
\end{equation}
\end{widetext}
with $\lambda_{\rho}=1.260$ fixed by demanding the proper normalization of the rho distribution amplitude (DA) for an on-shell rho of mass  
$m_\rho$
, as detailed in Appendix \ref{rho_LFWF}.
In other words, $\lambda_{\rho}=1.260$ reproduces the rho electroweak decay constant $f_\rho$. 
By setting the pseudoparticle (instanton) size $\rho=0.313$ fm, and a constituent mass $M=398.17$ MeV, the rho mass is found to be  $m_\rho=791$ MeV, and the normalization constants in (\ref{LF_T}-\ref{LF_L}) are
\begin{align}
   C_{\rho_L}=C_{\rho_T}=C_{\rho}&=1.9888
\end{align}
It is important to note that to recover full covariance on the light front
for the rho meson, terms of order $$p^--\frac{k_\perp^2+M^2}{2x\bar{x}p^+}\sim\mathcal{O}(n_{I+A}^2)$$ need to be retained in the light front analysis~\cite{Liu:2023fpj}.  Note that  Ward identity is readily enforced,
\begin{equation}
    p_\mu\bar{u}_{s_1}(k_1)\gamma^\mu v_{s_2}(k_2)=0
\end{equation}

Insering (\ref{LLX}-\ref{TTX}) into (\ref{tmd_rho}) and reducing, yield the leading twist TMDs in the ILM
\begin{widetext}
\begin{equation}
\begin{aligned}
f^{q/\rho}_1(x,k_\perp)=&\frac{1}{3}\frac{C^2_{\rho}}{(2\pi)^3}\left[\left(\frac{2}{m^2_\rho}+\frac{8x\bar{x}(k_\perp^2+M^2)}{(x\bar{x}m_{\rho}^2-k_\perp^2-M^2)^2}\right)\mathcal{F}^2\left(\frac{k_\perp}{\lambda_{\rho}\sqrt{x\bar x}}\right)\right]\\
&+\frac{2}{3}\frac{C^2_{\rho}}{(2\pi)^3}\left[\frac{2\left(k^2_\perp+M^2-2x\bar{x}k^2_\perp\right)}{(x\bar{x}m_{\rho}^2-k_\perp^2-M^2)^2}\mathcal{F}^2\left(\frac{k_\perp}{\lambda_{\rho}\sqrt{x\bar x}}\right)\right]
\end{aligned}
\label{F1QR}
\end{equation}

\begin{equation}
\begin{aligned}
f^{q/\rho}_{1LL}(x,k_\perp)=&\frac{C^2_{\rho}}{(2\pi)^3}\left[\left(\frac{2}{m^2_\rho}+\frac{8x\bar{x}(k_\perp^2+M^2)}{(x\bar{x}m_{\rho}^2-k_\perp^2-M^2)^2}\right)\mathcal{F}^2\left(\frac{k_\perp}{\lambda_{\rho}\sqrt{x\bar x}}\right)\right]\\
&-\frac{C^2_{\rho}}{(2\pi)^3}\left[\frac{2\left(k^2_\perp+M^2-2x\bar{x}k^2_\perp\right)}{(x\bar{x}m_{\rho}^2-k_\perp^2-M^2)^2}\mathcal{F}^2\left(\frac{k_\perp}{\lambda_{\rho}\sqrt{x\bar x}}\right)\right]
\end{aligned}
\label{F1LLQR}
\end{equation}

\begin{equation}
\begin{aligned}
f^{q/\rho}_{1LT}(x,k_\perp)=&\frac{C^2_{\rho}}{(2\pi)^3}\left[\frac{2(x-\bar{x})(x\bar{x}m^2_\rho+k_\perp^2+M^2)}{(x\bar{x}m_{\rho}^2-k_\perp^2-M^2)^2}\mathcal{F}^2\left(\frac{k_\perp}{\lambda_{\rho}\sqrt{x\bar x}}\right)\right]
\end{aligned}
\label{F1LTQR}
\end{equation}

\begin{equation}
\begin{aligned}
f^{q/\rho}_{1TT}(x,k_\perp)=&\frac{C^2_{\rho}}{(2\pi)^3}\left[\frac{4x\bar{x}m^2_\rho}{(x\bar{x}m_{\rho}^2-k_\perp^2-M^2)^2}\mathcal{F}^2\left(\frac{k_\perp}{\lambda_{\rho}\sqrt{x\bar x}}\right)\right]
\end{aligned}
\label{F1TTQR}
\end{equation}

\begin{equation}
\begin{aligned}
    &g^{q/\rho}_{1}(x,k_\perp)=\frac{C^2_{\rho}}{(2\pi)^3}\frac{4\left((x-\bar{x})k_\perp^2+M^2\right)}{\left(x\bar{x}m^2_\rho-k^2_\perp-M^2\right)^2}\mathcal{F}^2\left(\frac{k_\perp}{\lambda_{\rho}\sqrt{x\bar x}}\right)
\end{aligned}
\label{G1QR}
\end{equation}

\begin{equation}
\begin{aligned}
    g^{q/\rho}_{1T}(x,k_\perp)=\frac{C^2_{\rho}}{(2\pi)^3}\frac{2(x\bar{x}m^2_\rho+k_\perp^2+M^2)}{\left(x\bar{x}m^2_\rho-k^2_\perp-M^2\right)^2}\mathcal{F}^2\left(\frac{k_\perp}{\lambda_{\rho}\sqrt{x\bar x}}\right)
\end{aligned}
\label{G1TQR}
\end{equation}

\begin{equation}
\begin{aligned}
    h^{q/\rho}_{1}(x,k_\perp)=
        \frac{C^2_{\rho}}{(2\pi)^3}\frac{M}{m_\rho}\frac{2(x\bar{x}m^2_\rho+k_\perp^2+M^2)}{\left(x\bar{x}m^2_\rho-k^2_\perp-M^2\right)^2}\mathcal{F}^2\left(\frac{k_\perp}{\lambda_{\rho}\sqrt{x\bar x}}\right)
\end{aligned}
\label{H1QR}
\end{equation}

\begin{equation}
\begin{aligned}
    &h^{q/\rho}_{1L}(x,k_\perp)=
    -\frac{C^2_{\rho}}{(2\pi)^3}\frac{4\bar{x}Mm_\rho}{\left(x\bar{x}m^2_\rho-k^2_\perp-M^2\right)^2}\mathcal{F}^2\left(\frac{k_\perp}{\lambda_{\rho}\sqrt{x\bar x}}\right)
\end{aligned}
\label{H1LQR}
\end{equation}
\end{widetext}
\begin{equation}
\begin{aligned}
    &h^{q/\rho}_{1T}(x,k_\perp)=0
\end{aligned}
\end{equation}
We note the relationship 
\begin{equation}
g^{q/\rho}_{1T}(x,k_\perp)=\frac{m_\rho}{M}h^{q/\rho}_{1}(x,k_\perp)    
\end{equation}
which  is in agreement with the relationship noted in~\cite{Ninomiya:2017ggn}.
For comparison with lattice results and other works, we define the mean transverse momentum of a specific TMD function $F^{q/\rho}$ as
\begin{equation}
\label{kt_moment}
    \langle k_\perp \rangle = \frac{\int dx \, d^2k_\perp \, |k_\perp| F^{q/\rho}(x, k_\perp)}{\int dx \, d^2k_\perp \, F^{q/\rho}(x, k_\perp)}.
\end{equation}
The result is shown in Table \ref{tab:kt_values}.

\begin{table*}
    \centering
    \begin{tabular}{|c|c|c|c|c|c|c|c|}
        \hline
        & $\langle k_\perp \rangle_{\text{NJL}}$ 
        & $\langle k_\perp \rangle_{\text{LFHM}}$ 
        & $\langle k_\perp \rangle_{\text{LFQM}}$ 
        & $\langle k_\perp \rangle_{\text{BSE}}$ 
        & $\langle k_\perp \rangle_{\text{ILM}}$ \\
        \hline
        $f_1$    & 0.32 & 0.238 & 0.328 & 0.399 & 0.0810\\
        $g_{1}$ & 0.08 & 0.204 & 0.269 & 0.318 & 0.0751\\
        $g_{1T}$ & 0.34 & 0.229 & 0.269 & 0.358 & 0.0785\\
        $h_1$    & 0.34 & 0.229 & 0.307 & 0.367 & 0.0785\\
        $h_{1L}$ & 0.33 & 0.204 & 0.269 & 0.368 & 0.0751\\
        $h_{1T}$ & ---  & ---   & 0.237 & 0.365 & ---\\
        $f_{1TT}$ & 0.32 & 0.211 & ---   & 0.338 & 0.0734\\
        \hline
    \end{tabular}
    \caption{Table of the $k_\perp$-moment of TMDs defined in \eqref{kt_moment}. Lines
in the blank indicate the corresponding TMDs are vanishing.
All units are given in GeV. Our results are compared to Nambu–Jona-Lasinio (NJL) model \cite{Ninomiya:2017ggn}, light-front holographic model (LFHM) and light-front quark model (LFQM) \cite{Kaur:2020emh} and Bethe-Sapeter equation (BSE) \cite{Shi:2022erw}}
    \label{tab:kt_values}
\end{table*}

\subsection{TMDs at low resolution}

In Fig.~\ref{fig:rho_TMDf1}, we show the rho TMD $f^{q/\rho}_1$ (\ref{F1QR}) (left column) 
 and the rho TMD $f^{q/\rho}_{1LT}$ (\ref{F1LTQR}) (right column), in the 2-Fock space approximation, calculated by the ILM at a renormalization scale, or resolution, $\mu=1/\rho$ with $\rho=0.313\,\rm  fm$. The rho LFWF parameters are $C_{\rho}=1.9888$, $m_\rho=791.0$ MeV, $M=398.17$ MeV. 
(\ref{fig:rhof1_contour}, \ref{fig:rhof1LT_contour}) are the 2D density plots in $x, k_\perp$, and (\ref{fig:rhof1_3d}, \ref{fig:rhof1LT_3d}) their  3D visualization. 
The support of the TMDs in longitudinal momentum is dominant at $x\approx \frac 12$, with a width  $\sqrt{\langle k^2_\perp\rangle}\approx 0.1\,\rm GeV$. 
The profiles of the TMDs (\ref{F1QR}) and (\ref{F1LTQR}), in parton $x$  for $k_\perp=0.2, 0.3, 0.4\, \rm GeV$ are shown in (\ref{fig:rhof1_x}, \ref{fig:rhof1LT_x}), and the corresponding profiles in $k_\perp$ for $x=0.1, 0.2, 0.3$ are shown in (\ref{fig:rhof1_k}, \ref{fig:rhof1LT_k}), respectively.
For fixed $k_\perp$, the distribution in parton-$x$ of $f_1^{q/\rho}$ is symmetric, while that of  $f_{1LT}^{q/\rho}$ is antisymmetric. Both distributions in $k_\perp$ for fixed parton-$x$, are Gaussian-like. This final feature arises from the profile of the quark zero modes in the ILM. 


In Fig.~\ref{fig:rho_TMDf1TT} we show the rho TMD $f_{1LL}^{q/\rho}$ (\ref{F1LLQR}) (left column) 
 and the rho TMD $f_{1TT}^{q/\rho}$ (\ref{F1TTQR}) (right column), in the 2-Fock space approximation, in the ILM at a resolution $\mu=1/\rho$ with $\rho=0.313\,\rm fm$. The rho LFWF parameters are $C_{\rho}=1.9888$, $m_\rho=791.0$ MeV, $M=398.17$ MeV.
(\ref{fig:rhof1LL_contour},\ref{fig:rhof1TT_contour}) are the 2D density plots in $x, k_\perp$, and (\ref{fig:rhof1LL_3d},\ref{fig:rhof1TT_3d}) their  3D  visualization. 
The support of the TMDs in longitudinal momentum is dominant at $x\approx \frac 12$, with a width  $\sqrt{\langle k^2_\perp\rangle}\approx 0.1\,\rm GeV$. 
The profiles of the TMDs (\ref{F1LLQR}) and (\ref{F1TTQR}), in parton-$x$  for $k_\perp=0.2, 0.3, 0.4\, \rm GeV$ are shown in (\ref{fig:rhof1LL_x},\ref{fig:rhof1TT_x}), and the corresponding profiles in $k_\perp$ for $x=0.1, 0.2, 0.3$ are shown in (\ref{fig:rhof1LL_k},\ref{fig:rhof1TT_k}), respectively.
For fixed $k_\perp$, both distributions in parton-$x$ of $f_{LL}^{q/\rho}$ and  $f_{1TT}^{q/\rho}$ are  symmetric, with the former displaying 2 zeros, an indication of a possible P-wave admixture.
Both distributions in $k_\perp$ for fixed parton-$x$, are Gaussian-like, the feature introduced by the profile of the instanton zero modes.



In Fig.~\ref{fig:rho_TMDg1} we show the rho TMD $g_1^{q/\rho}$ (\ref{G1QR}) (left column) 
 and the rho TMD $g_{1T}^{q/\rho}$ (\ref{G1TQR}) (right column), in the 2-Fock space approximation, in the ILM at a resolution $\mu=1/\rho$ with $\rho=0.313\,\rm  fm$. The rho LFWF parameters are $C_{\rho}=1.9888$, $m_\rho=791.0$ MeV, $M=398.17$ MeV.
(\ref{fig:rhog1_contour},\ref{fig:rhog1T_contour}) are the 2D density plots in $x, k_\perp$, and (\ref{fig:rhog1_3d},\ref{fig:rhog1T_3d}) their 3D  visualization. 
The support of the TMDs in longitudinal momentum is dominant at $x\approx \frac 12$, with a wider spread in  $\sqrt{\langle k^2_\perp\rangle}\approx 0.1\,\rm GeV$. 
The profiles of the TMDs (\ref{G1QR}) and (\ref{G1TQR}), in parton-$x$  for $k_\perp=0.2, 0.3, 0.4\, \rm GeV$ are shown in (\ref{fig:rhog1_x},\ref{fig:rhog1T_x}), and the corresponding profiles in $k_\perp$ for $x=0.1, 0.2, 0.3$ are shown in (\ref{fig:rhog1_k},\ref{fig:rhog1T_k}), respectively.
For fixed $k_\perp$, both distributions in parton-$x$ of $g_{1}^{q/\rho}$ and  $g_{1T}^{q/\rho}$ are  symmetric.
Both distributions in $k_\perp$ for fixed parton-$x$, are Gaussian-like, due to the profile of the instanton zero modes.


In Fig.~\ref{fig:rho_TMDh1} we show the rho TMD $h_1^{q/\rho}$ (\ref{H1QR}) (left column) and the rho TMD $h_{1L}^{q/\rho}$ (\ref{H1LQR}) (right column), calculated by the 2-Fock space approximation, in the ILM at a resolution $\mu=1/\rho$ with $\rho=0.313\,\rm  fm$. The rho LFWF parameters are $C_{\rho}=1.9888$, $m_\rho=791.0$ MeV, $M=398.17$ MeV.
(\ref{fig:rhoh1_contour},\ref{fig:rhoh1L_contour}) are the 2D density plots in $x, k_\perp$, and (\ref{fig:rhoh1_3d},\ref{fig:rhoh1L_3d}) their 3D visualization. The support of the TMDs in longitudinal momentum is dominant at $x\approx \frac 12$, with a wider spread in  $\sqrt{\langle k^2_\perp\rangle}\approx 0.1\,\rm GeV$. 
The profiles of the TMDs (\ref{H1QR}) and (\ref{H1LQR}), in parton-$x$  for $k_\perp=0.2, 0.3, 0.4\, \rm GeV$ are shown in (\ref{fig:rhoh1_x},\ref{fig:rhoh1L_x}), and the corresponding profiles in $k_\perp$ for $x=0.1, 0.2, 0.3$ are shown in (\ref{fig:rhoh1_k},\ref{fig:rhoh1L_k}), respectively.
For fixed $k_\perp$, both distributions in parton-$x$ of $h_{1}^{q/\rho}$ and  $h_{1L}^{q/\rho}$ are  symmetric.
Both distributions in $k_\perp$ for fixed parton-$x$, are Gaussian-like due to the profile of the zero modes in the ILM.

\subsection{Rho PDFs from TMDs}
The integration over the transverse momentum $k_\perp$ dependence of the TMDs
yields the PDFs. As we noted in the introduction and for  vector mesons, we have four twist-2 PDFs $f_1^{q/\rho}(x)$, $f_{1LL}^{q/\rho}(x)$, $g_1^{q/\rho}(x)$, and $h_1^{q/\rho}(x)$ given by \cite{Ninomiya:2017ggn}
\begin{equation}
\label{pdf1}
    \int d^2k_\perp q_\rho(x,k_\perp)=f_1^{q/\rho}(x)+S_{LL}f_{1LL}^{q/\rho}(x)
\end{equation}
\begin{equation}
\label{pdf2}
    \int d^2k_\perp \Delta q_\rho(x,k_\perp)=\lambda s_Lg_1^{q/\rho}(x)
\end{equation}
\begin{equation}
\label{pdf3}
    \int d^2k_\perp \delta q^\alpha_\rho(x,k_\perp)=\lambda s^\alpha_\perp h_1^{q/\rho}(x)
\end{equation}
with 
\begin{equation}
\begin{aligned}
&\left[f^{q/\rho}_1,~f^{q/\rho}_{1LL},~g^{q/\rho}_1,~h^{q/\rho}_1\right](x)\\
    =&\int d^2k_\perp\left[f^{q/\rho}_1,~f^{q/\rho}_{1LL},~g^{q/\rho}_1,~h^{q/\rho}_1\right](x,k_\perp)
\end{aligned}
\end{equation}
For a spin-1 meson, the quark and anti-quark  distributions are related by charge conjugation symmetry  
\begin{equation}
\begin{aligned}
&\left[f^{u/\rho^+}_1,~f^{u/\rho^+}_{1LL},~g^{u/\rho^+}_1,~h^{u/\rho^+}_1\right](x)\\
=&\left[f^{\bar{d}/\rho^+}_1,~f^{\bar{d}/\rho^+}_{1LL},~g^{\bar{d}/\rho^+}_1,~h^{\bar{d}/\rho^+}_1\right](1-x)
\end{aligned}
\end{equation}
and
\begin{equation}
\begin{aligned}
&\left[f^{u/\rho^+}_1,~f^{u/\rho^+}_{1LL},~g^{u/\rho^+}_1,~h^{u/\rho^+}_1\right](x)\\
=&\left[f^{\bar{u}/\rho^-}_1,~f^{\bar{u}/\rho^-}_{1LL},~g^{\bar{u}/\rho^-}_1,~h^{\bar{u}/\rho^-}_1\right](x)
\end{aligned}
\end{equation}

For spin-1 mesons there are four PDFs. Compared to the three nucleon PDFs, the new parton distribution $f^{q/\rho}_{1LL}$ was noted originally by Hoodbhoy, Jaffe and Manohar~\cite{Hoodbhoy:1988am}. This new PDF, also referred to as  $b^q_1$ structure function, represents the difference of unpolarized quark distributions in a longitudinally polarized spin-1 meson with spin projection $\lambda=0$ and $\lambda=\pm1$,
\begin{equation}
\begin{aligned}
    &\frac{1}2 f^{q/\rho}_{1LL}(x)\equiv b^{q/\rho}_1(x) \\
    =& \frac{1}{4} \left[ 2q_\rho^{(\lambda=0)}(x) - q_\rho^{(\lambda=1)}(x) - q_\rho^{(\lambda=-1)}(x) \right]
\end{aligned}
\end{equation}
This PDF is sensitive to the orbital motion of the constituents, and may also be carrying information on QCD exotica~\cite{Miller:2013hla,Liuti:2014dda}.

\subsection{Rho sum rules}
The standard parton distribution sum rule guarantees
\begin{align}
\label{g_sum}
    \int_0^1dx f_1^{q/\rho}(x)=1
\end{align}
The first moment of the unpolarized parton distribution $f_{1}^{q/\rho}$,
is about the longitudinal momentum carried by the quarks. In Table~\ref{tab:x_values} we compare our results with those of other effective models,
and selected lattice results.

For spin-1 mesons, there is an additional sum rule~\cite{Close:1990zw,Efremov:1982high} 
\begin{align}
\label{fLL_sum}
     \int_0^1dx f_{1LL}^{q/\rho}(x)=0
\end{align}
which reflects on the fact that the valence quark number is 
independent of the hadron spin state.
It was shown  in~\cite{Umnikov:1996qv}
that Lorentz covariance is important for the enforcement of this sum rule,
even for  non-relativistic constituents. The first moment of $f_{1LL}^{q/\rho}$ is compared to the lattice result from~\cite{Zhang:2024plq} at the resolution  $\approx 2.4$ GeV in Table~\ref{tab:xLL_values}. If the quarks are in a relative S-wave state in the infinite momentum frame, its first moment is zero.  The lattice results are evidence for an orbital contribution to the quark content~\cite{Zhang:2024plq}.  The null result for our analysis in the ILM,
suggests that a higher P-wave Fock contribution to the rho meson is needed, to account for the missing orbital contribution.


\begin{table*}
    \centering
    \begin{tabular}{|c|c|c|c|c|c|c|c|}
        \hline
        & NJL \cite{Ninomiya:2017ggn}
        & Lattice \cite{Zhang:2024plq}
        & ILM ($\mu\sim1/\rho$)
        & ILM ($\mu=2.4$ GeV)\\
        \hline
        $ \int dx xf_1^{q/\rho}(x)$  & 0.5 & 0.334(21) & 0.5 & 0.201 \\
        \hline
    \end{tabular}
    \caption{Our instanton estimation on the first moment of unpolarized PDF $f^{q/\rho}_1$ is compared to NJL model \cite{Ninomiya:2017ggn}, and lattice calculation on the rho meson structure function with renormalization $\approx 2.4$ GeV. We also evolve our result to $2.4$ GeV by Dokshitzer–Gribov–Lipatov–Altarelli–Parisi (DGLAP) equation}
    \label{tab:x_values}
\end{table*}

\begin{table*}
    \centering
    \begin{tabular}{|c|c|c|c|c|c|c|c|}
        \hline
        & NJL \cite{Ninomiya:2017ggn}
        & Lattice \cite{Zhang:2024plq}
        & ILM ($\mu\sim1/\rho$)
        & ILM ($\mu=2.4$ GeV)\\
        \hline
        $\int_0^1dx xf_{1LL}^{q/\rho}(x)$  & 0 & 0.29(23) & 0 & 0 \\
        \hline
    \end{tabular}
    \caption{Our instanton estimation on the first moment of tensor polarized PDF $f^{q/\rho}_{1LL}$ is compared to NJL model \cite{Ninomiya:2017ggn}, and lattice calculation on the rho meson structure function with renormalization $\approx 2.4$ GeV. We also evolve our result to $2.4$ GeV by DGLAP equation}
    \label{tab:xLL_values}
\end{table*}

The quark spin sum (axial charge) for each quark in the meson state is defined as
\begin{equation}
\label{h_sum}
    \Delta q=\int_0^1dx g_1^{q/\rho}(x)
\end{equation}
The comparison of our ILM estimate to other models  and selected lattice results, is shown in Table~\ref{tab:axial_values}. The results in the ILM show that the total valence quark and antiquark contribution to the spin of the $\rho$ meson, is $96.7\%$. 
\begin{table}
    \centering
    \begin{tabular}{|c|c|c|c|c|c|c|c|}
        \hline
        & NJL \cite{Ninomiya:2017ggn}
        & BSE \cite{Shi:2022erw}
        & Lattice \cite{Zhang:2024plq}
        & ILM \\
        \hline
        $\Delta q$    & 0.56 & 0.67 & 0.570(32) & 0.967\\
        \hline
    \end{tabular}
    \caption{Table of the quark spin sum defined in \eqref{g_sum}. 
 Our instanton estimation is compared to NJL model \cite{Ninomiya:2017ggn}, Bethe-Salpeter equation (BSE) \cite{Shi:2022erw}, and Lattice calculation on the rho meson structure function with renormalization $\approx 2.4$ GeV.}
    \label{tab:axial_values}
\end{table}

The tensor charge for each quark in the rho meson state is defined as
\begin{equation}
    \delta q=\int_0^1dx h_1^{q/\rho}(x)
\end{equation}
The tensor charge is renormalization-scale dependent~\cite{He:1994gz}. The 1-loop RG evolution reads
\begin{equation}
    \delta q(\mu) = \left( \frac{\alpha_s(\mu)}{\alpha_s(\mu_0)} \right)^{\frac{C_F}{\beta_0}} \delta q(\mu_0),
\end{equation}
The comparison of our ILM estimation to other model is shown in Table~\ref{tab:tensor_values}. Our instanton estimation shows the tensor charge of the $\rho$ meson is $\delta q=0.9807$ at original instanton scale $1/\rho\simeq630$ MeV and $\delta q=0.704$ at $2$ GeV
\begin{table*}
    \centering
    \begin{tabular}{|c|c|c|c|c|c|c|c|}
        \hline
        & NJL \cite{Ninomiya:2017ggn}
        & Lattice \cite{Zhang:2024plq}
        & BSE \cite{Shi:2022erw}
        & ILM ($\mu\sim1/\rho$)
        & ILM ($\mu=2$ GeV) \\
        \hline
        $\delta q$    & 0.94 & 0.803 & 0.79 & 0.981 & 0.704\\
        \hline
    \end{tabular}
    \caption{Table of the tensor charge defined in \eqref{h_sum}. 
 Our instanton estimation is compared to NJL model \cite{Ninomiya:2017ggn} and Bethe-Sapeter equation (BSE) \cite{Shi:2022erw}}
    \label{tab:tensor_values}
\end{table*}

\begin{figure}
    \centering
\subfloat[]{\includegraphics[width=1\linewidth]{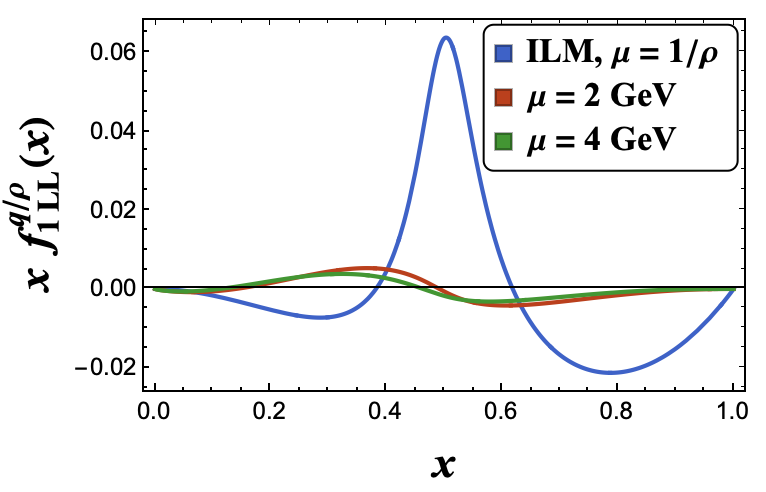}}
\hfill
\subfloat[]{\includegraphics[width=1\linewidth]{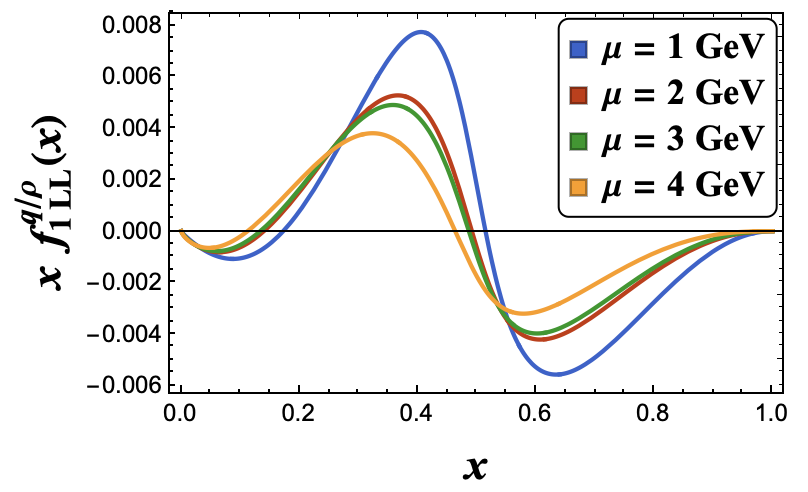}}
    \caption{The rho parton distribution $f_{1LL}^{q/\rho}$  at  $\mu\simeq1/\rho$ (a), and after  DGLAP evolution (b). }
    \label{fig:xpdf_rho_h1}
\end{figure}

\begin{figure*}
    \centering
\subfloat[]{\includegraphics[width=.33\linewidth]{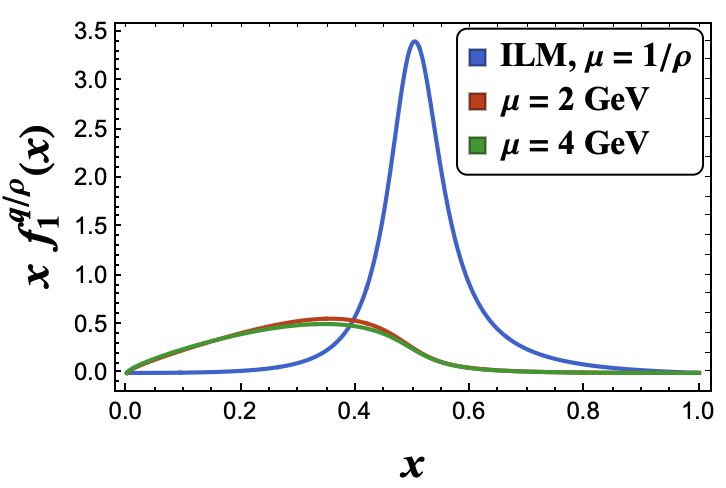}}
\hfill
\subfloat[]{\includegraphics[width=.33\linewidth]{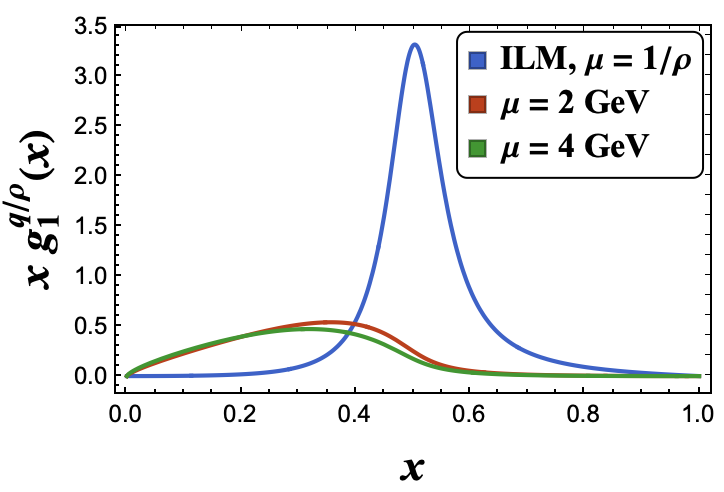}}
\hfill
\subfloat[]{\includegraphics[width=.33\linewidth]{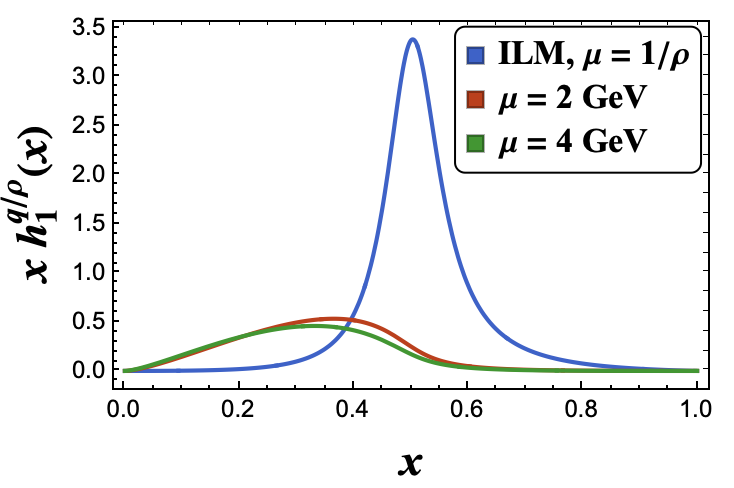}}
\hfill
\subfloat[]{\includegraphics[width=.33\linewidth]{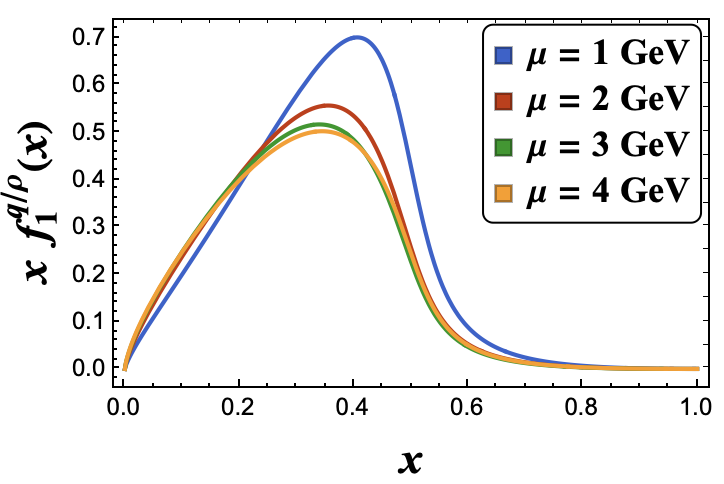}}
\hfill
\subfloat[]{\includegraphics[width=.33\linewidth]{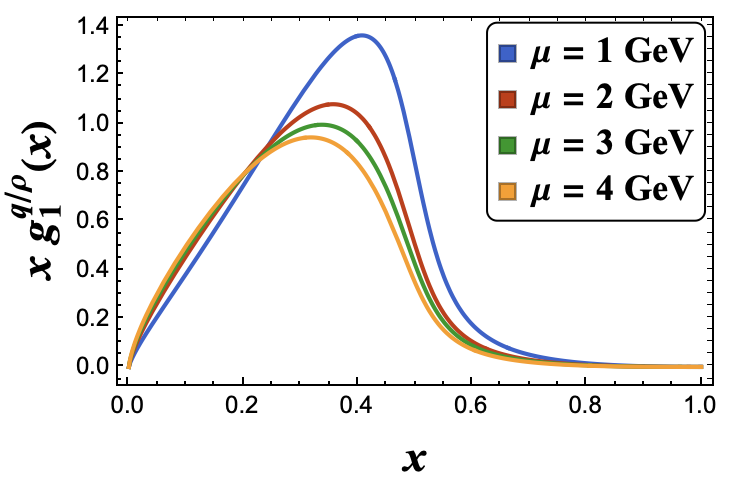}}
\hfill
\subfloat[]{\includegraphics[width=.33\linewidth]{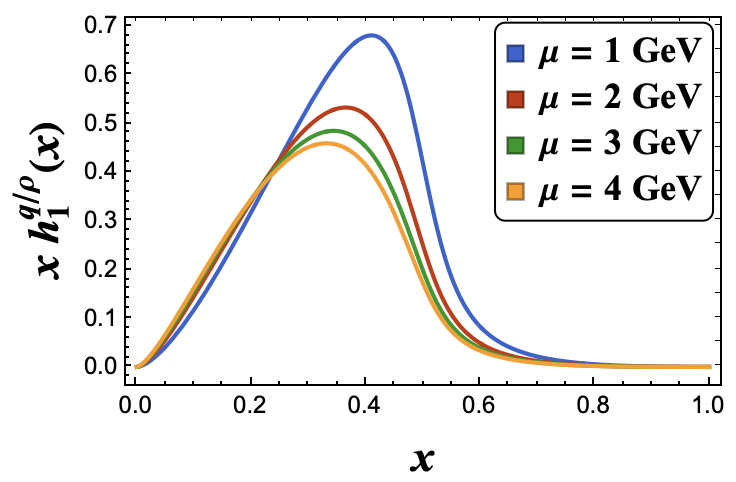}}
    \caption{The rho parton distributions at $\mu\simeq1/\rho$:  (a) $f_1^{q/\rho}$, (b) $g_1^{q/\rho}$, and (c) $h_1^{q/\rho}$ compared to their DGLAP evolved counterparts  (d), (e), (f). }
    \label{fig:xpdf_rho}
\end{figure*}

In Fig.~\ref{fig:xpdf_rho_h1} we show 
the rho meson PDF $f_{1LL}^{q/\rho}$ at the resolution $\mu\simeq1/\rho$ in the ILM, and their evolved version at different resolutions $\mu=1,2,3,4 \, \rm GeV$ (b). These rho PDFS are obtained before the soft subtraction to be discussed below. 
The rho meson PDFs  $f_1^{q/\rho}$ \eqref{pdf1}, $g_1^{q/\rho}$ \eqref{pdf2}, and $h_1^{q/\rho}$ \eqref{pdf3} are shown in Fig.~\ref{fig:xpdf_rho} at the resolution $\mu\simeq1/\rho$ (a) $f_1^{q/\rho}$, (b) $g_1^{q/\rho}$, and (c) $h_1^{q/\rho}$ compared to the results evolved to $2$ and $4$ GeV by DGLAP. Their other evolved results 
at different resolutions $\mu=1,2,3,4 \, \rm GeV$, are  shown in (d), (e), (f) respectively. Overall, there is substantial shift of the evolved PDFs towards low-$x$.

\section{Soft subtraction and TMD evolution}
\label{SECIII}
\subsection{Soft subtracted TMD distributions}
The definition of TMD distributions as twist-2 matrix matrix elements in \eqref{tmd}, accounts for the transverse momentum dependent structure of a parton inside a hadron in the entire range of kinematics. 
In the following, we will denote them by $\tilde{F}^{q/\rho}(x,b_\perp,\mu,y_q-y_n)$, where the rapidity arises from the stapled Wilson lines. However, in TMD factorization, the definition in \eqref{tmd}, which is refered to unsubtracted TMD functions, often leads to an ambiguous overlap between the collinear and soft kinematic region. This leads to lightcone divergences that appear in the light-like Wilson lines, regulated by $y_n$, in both the TMD distributions and soft factor. This is because in the real world, these gluon exchanges are not distinguishable. Therefore, to disentangle their contributions in the factorization, we introduce rapidity regularization by implementing a soft subtraction.

For a generic unsubtracted rho meson TMD say $\tilde{F}^{q/\rho}(x,b_\perp,\mu,y_q-y_n)$, defined in \eqref{tmd}, the soft subtracted TMDs in $b_\perp$ follow through
\begin{equation}
\begin{aligned}
\label{eq:tmd_b}
    \tilde{F}^{q/\rho}(x,b_\perp;\mu,\zeta)=&
    \frac{\tilde{F}^{q/\rho}(x,b_\perp,\mu,y_q-y_n)}{S(b_\perp,\mu,y-y_n)}
\end{aligned}
\end{equation}
where $y_q=\frac12\ln(k^+/k^-)$ is the hadron rapidity and $y_n=\frac12\ln(n^+/n^-)$ is the rapidity for the near-light-cone direction $n$ of Wilson line in the hadronic matrix element defintion \eqref{tmd}. This direction $n$ usually refers to the moving direction of the other hadron in the process. The rapidity scale 
$$\zeta=2(k^+)^2e^{-y}$$
represents the energy of the hadron with a rapidity regulator $y$, distinguishing the energy between two involving hadrons in the TMD factorization process, and $\zeta_\rho\zeta_h=Q^4$ where $\zeta_h$ is the rapidity scale of the other involved hadron $h$. 
The soft factor $S(b_\perp,\mu,y_n-y_{\bar n})$ is defined as \cite{Collins:2011zzd,Ji:2004wu,Liu:2024sqj}
\begin{equation}
\begin{aligned}
    &S(b_\perp,\mu,y_n-y_{\bar n})=\\
    &\frac1{N_c}\Tr\langle W_{\pm\bar n}(0)W^\dagger_{\pm n}(0)W_{\pm n}(b_\perp)W^\dagger_{\pm\bar n}(b_\perp)\rangle
\end{aligned}
\end{equation}



\subsection{TMD evolution}
For any kind of leading-twist TMD functions with soft subtraction $\tilde{F}^{q/\rho}$ in $b_\perp$ space, the CSS renormalization group equations read \cite{Boer:2015ala,Collins:2014loa}
\begin{widetext}
\begin{align}
    \frac{d}{d\ln\sqrt{\zeta}} \ln \tilde{F}^{q/\rho}(x, b_\perp; \mu,\zeta)=&K_{\rm CS}(b_\perp,\mu)\nonumber\\
    \frac{d}{d\ln\mu} \ln\tilde{F}^{q/\rho}(x, b_\perp; \mu,\zeta)=& \gamma_F (\alpha_s(\mu))- \Gamma_{\mathrm{cusp}}(\alpha_s(\mu)) \ln\left(\frac{\zeta}{\mu^2}\right)
\end{align}
\end{widetext}
with the Collin-Sober kernel $K_{\rm CS}$, also known as the rapidity anomalous dimension that governs the rapidity $\zeta$ scale evolution. The anomalous dimension $\gamma_F$ depends on the quark spin sources  $\gamma^+,\gamma^+\gamma^5, i\sigma^{\alpha+}\gamma^5$. The perturbative expansions of the anomalous dimension functions $\Gamma_{\rm cusp}$ and $\gamma_F$, are given in Appendix~\ref{Appx:cusp}.

\begin{figure}
    \centering
    \includegraphics[width=1\linewidth]{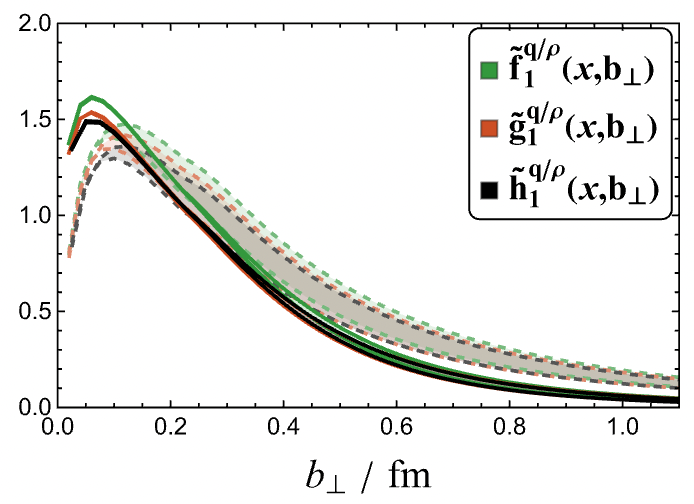}
    \caption{Soft subtracted unpolarized $\tilde{f}_1^{q/\rho}$ (green), spin-dependent TMD distributions $\tilde{g}_1^{q/\rho}$ (red)  and $\tilde{h}_1^{q/\rho}$ (black) at two different high resolution $\mu=2$ GeV (dashed lines) and $\mu=10$ GeV (solid lines) using \eqref{INTER}, with the bands corresponding to the fitting parameter $b_{\rm NP}=1.3-2.0$ GeV$^{-1}$ \cite{Liu:2024sqj}. }
    \label{fig:enter-labelTOT}
\end{figure}

\begin{figure*}
    \centering
\subfloat[]{\includegraphics[width=0.33\linewidth]{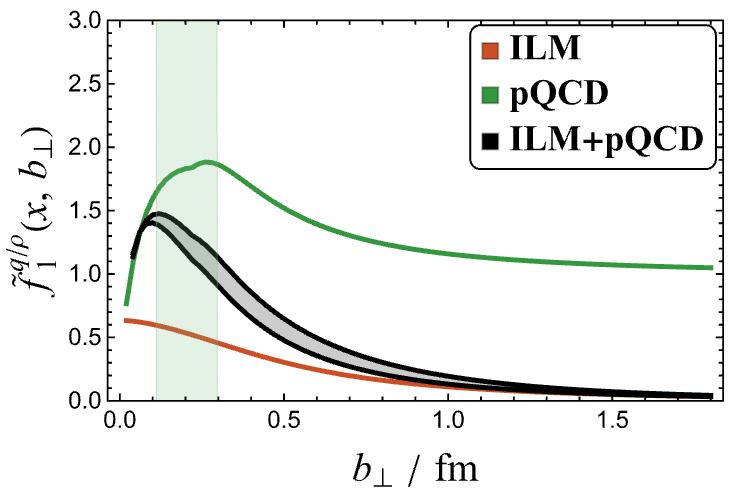}}
\hfill 
\subfloat[]{\includegraphics[width=0.33\linewidth]{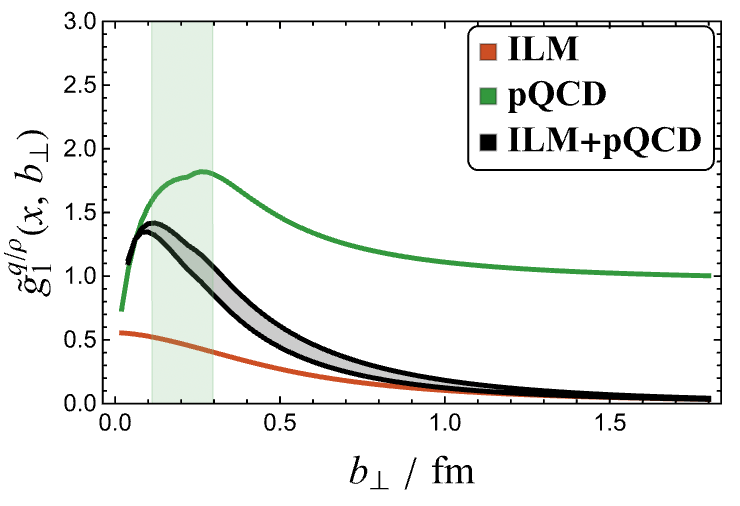}}
\hfill 
\subfloat[]{\includegraphics[width=0.33\linewidth]{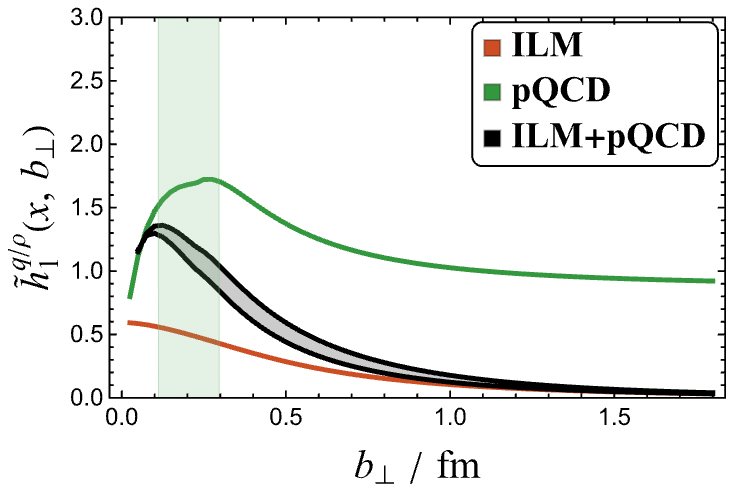}}
    \caption{
        The unpolarized TMD distributions (a) $\tilde{f}_1^{q/\rho}$, and spin-dependent TMDs (b) $\tilde{g}_1^{q/\rho}$ , (c) $\tilde{h}_1^{q/\rho}$  
    are shown at the resolution $\mu=2\,\rm GeV$ for also $x=0.3$ with black bands corresponding to the interpolating parameter $b_{\rm NP}=1.3 - 2.0$ GeV$^{-1}$ \eqref{INTER}. The green interval corresponds to the perturbatively reliable region $2e^{-\gamma_E}/\mu<b_{\perp}<1.5$ GeV$^{-1}$.}
    \label{fig:enter-label}
\end{figure*}

The TMDs  involve two evolution scales $\mu$, $\zeta$, and one intrinsic transverse scale $b_\perp$. Their evolution is driven
by both a perturbative and non-perturbative kernels~\cite{Collins:2014jpa,Scimemi:2016ffw,DAlesio:2014mrz,Collins:1984kg,Rogers:2015sqa}. 
For simplicity, we can choose a simple  evolution scheme, with $\zeta=\mu^2$. In the  perturbative region $b_\perp\ll b_{\rm NP}$, the transverse scale of the TMD at high resolution $\mu\gg1/\rho$ is completely governed by the perturbative analysis, with the result
\begin{widetext}
\begin{equation}
\begin{aligned}
\label{convo}
    &\tilde{F}^{q/\rho}(x, b_\perp; \mu,\mu^2)\big|_{b\ll b_{\rm NP}}
    \simeq\sum_{q'=q, \bar q, g}\int_x^1 \frac{dx'}{x'}C^f_{q/q'}(x/x',b_\perp,\mu_b)f^{q'/\rho} (x',\mu_b)\\
    &\times \exp\left[- \int_{\mu_b}^{\mu} \frac{d\mu'}{\mu'} \left[ \Gamma_{\rm cusp}(\alpha_s(\mu'^2)) \ln \left( \frac{\mu^2}{\mu'^2} \right) - \gamma_F(\alpha_s(\mu'^2)) \right]+K^{(\rm pert)}_{\rm CS}(b_{\perp},\mu_b)\ln\left(\frac{\mu}{\mu_b}\right)\right]
\end{aligned}
\end{equation}
\end{widetext}
Here $f^{q/\rho} (x,\mu_b)=\int d^2k_\perp F^{q/\rho}(x,k_\perp)$ is the corresponding leading-twist collinear PDF, namely $f^{q/\rho}_1$, $f^{q/\rho}_{1LL}$, $g^{q/\rho}_1$, $h^{q/\rho}_1$, at $\mu_b$ scale defined as

\begin{equation}
\label{mu_b}
    \mu_b=2e^{-\gamma_E}/b_\perp
\end{equation}

The convolution is the operator product expansion (OPE), which describes the small-$b_\perp$ behavior of the TMDs in terms of the collinear PDFs $f^{q/\rho}$, convoluted with perturbative Wilson coefficients $C^f_{i/j}$ where $f$ denotes the type of the collinear PDF. The details are briefly given in Appendix \ref{Appx:cusp}.




In the non-perturbative region $b_\perp\gg b_{\rm NP}$, the instanton 
and anti-instanton fields contribute. The correponding TMD  reads
\begin{widetext}
\begin{equation}
\begin{aligned}
\label{F_np}
    \tilde{F}^{q/\rho}(x,b_\perp;\mu,\mu^2)\big|_{b_\perp\gg b_{\rm NP}}&\simeq~\tilde{F}^{q/\rho}(x,b_\perp;\mu_0,\mu_0^2) e^{K^{(\rm inst)}_{\rm CS}(b_\perp/\rho)\ln\sqrt{\frac{\mu^2}{\mu_0^2}}}\\
    &\times \exp\left[\int_{\mu_0}^{\mu} \frac{d\mu'}{\mu'}\left(\gamma_F(\alpha_s(\mu'^2))-\Gamma_{\rm cusp}(\alpha_s(\mu'^2)) \ln \left( \frac{\mu_0^2}{\mu'^2} \right) \right)\right]
\end{aligned}
\end{equation}
where the 
low resolution TMD in the ILM is defined by
\begin{equation}
\begin{aligned}
    \tilde{F}^{q/\rho}(x,b_\perp;\mu_0,\zeta_0)=&\left[\int d^2k_\perp e^{ik_\perp\cdot b_\perp}F_{(\rm inst)}^{q/\rho}(x,k_\perp)\right]e^{K^{(\rm inst)}_{\rm CS}(b_\perp/\rho)\ln\sqrt{\frac{\zeta_0}{1/\rho^2}}}
\end{aligned}
\end{equation}
\end{widetext}
The unsubstracted TMD at low resolution is composed of the overlap of the LFWFs $F_{(\rm inst)}^{q/\rho}(x,k_\perp)$ in~\eqref{tmd_rho} and a rapidity factor $\exp\left(K^{(\rm inst)}_{\rm CS}\ln\sqrt{\zeta_0/x^2m^2_\rho}\right)$ from the stapled Wilson line. Thus, the low energy unsubstracted TMD can be fully calculated in the instanton liquid model.

The evolution now is separated into a perturbative and non-perturbative region, where the cut-off separating the  two regions, is set as  $$b_{\rm NP}=\frac{2e^{-\gamma_E}}{\mu_0}=1.5~\mathrm{GeV}^{-1}$$ where $\mu_0\sim1/\rho$ is the non-perturbative scale fixed in the ILM.
The full TMD at high resolution $\mu\gg1/\rho$ interpolates between the large $b_\perp$ and small $b_\perp$, by optimal matching
the linear combination
\begin{widetext}
\begin{equation}
\label{INTER}
\tilde{F}^{q/\rho}(x, b_\perp; \mu,\mu^2)=\tilde{F}^{q/\rho}(x, b_\perp\ll b_{\rm NP}; \mu,\mu^2)e^{-b_\perp/b_{\rm NP}}+\tilde{F}^{q/\rho}(x, b_\perp\gg b_{\rm NP}; \mu,\mu^2)\left(1-e^{-b_\perp/b_{\rm NP}}\right)
\end{equation}
\end{widetext}
with $b_{\rm NP}$ the optimization parameter.

In Fig.~\ref{fig:enter-labelTOT} we show the unpolarized $\tilde{f}_1^{q/\rho}$ (green), spin-dependent $\tilde{g}_1^{q/\rho}$ (red)  and $\tilde{h}_1^{q/\rho}$ (black) TMDs at the resolution $\mu=2$ GeV (solid lines) compared to a higher resolution $\mu=10$ GeV (dashed lines). The results follow from~\eqref{INTER} with the matching parameter in the range $b_{\rm NP}=1.3-2.0$ GeV$^{-1}$ \cite{Liu:2024sqj}. In Fig.~\ref{fig:enter-label} we show the separate results for the  corresponding TMDs (a) $\tilde{f}_1^{q/\rho}$ , (b) $\tilde{g}_1^{q/\rho}$, and (c) $\tilde{h}_1^{q/\rho}$ at $x=0.3$, with the pQCD results (green), the ILM (red) and combined result (black) with a band corresponding to the matching parameter in the range $b_{\rm NP}=1.3-2.0$ GeV$^{-1}$ at high resolution $\mu= 2\,\rm GeV$ with $\zeta=\mu^2$. Only the small-$b_\perp$ perturbative contribution inside green band ($b_{\perp}=0.56 - 1.5$ GeV$^{-1}$) is perturbatively reliable. The smaller region ($b_\perp\leq1/\mu$) suffers the sensitivity to the UV regularization scheme and uncertainty from the large logarithm. The phenomenological prescriptions to this region have been proposed in \cite{Bacchetta:2017gcc,Bacchetta:2022awv,Rogers:2024cci,Boer:2014tka} (and references therein). On the other hand, large $b_\perp$ TMD is dominated by ILM. The large $b_\perp$ of perturbative contribution in Fig.~\ref{fig:enter-label} is simply modified by $b_*$-prescription \cite{Collins:2016hqq} such that the saturation occurs to avoid the Landau pole $\Lambda_{\rm QCD}$. 


\section{Conclusions}
\label{SECIV}
We have developed an analysis of the  leading-twist TMDs of the spin-1 rho meson, using the ILM of the QCD vacuum. By making use of the most general 
spin quantization, we derived all nine T-even TMDs for the rho meson, and
their corresponding reductions to PDFs. The analysis uses a leading Fock approximation for the longitudinal and transverse rho meson light front wavefunctions. $O(3)$ symmetry is enforced through proper counting rule of instanton density $n_{I+A}$
in the ILM, at low resolution. Lorentz covariance  is dynamically restored, with the momentum sum rules for the unpolarized and tensor polarized PDFs satisfied in the ILM.

Before evolution and at the ILM resolution $\mu=1/\rho$, the
$h_{1T}^{q/\rho}$ rho TMD  is found to be null in the leading Fock approximation. The  remaining and non-vanishing T-even rho TMDs show a Gaussian-like behavior in $k_\perp$ for fixed parton-$x$. This  feature follows from a profiling of the zero modes in the ILM, and carries to the rest of the TMDs.
The $f_{1TT}^{q/\rho}$ rho TMD is found to be antisymmetric with respect to the $x=\frac 12$-line, while the $f_{1LL}^{q/\rho}$ vanishes around two symmetric lines about $x=\frac 12$, a possible sensitivity to a small P-wave contribution in the spin-1 state,  in the 2-Fock space approximation. At low resolution, the rho tensor charge  is about 1, in agreement with the constituent quark model, or NJL.

Following our recent analysis of the pion and kaon TMDs~\cite{Liu:2025mbl}, we have suggested that the rho TMDs in the ILM at low resolution, should be supplemented with the pertinent CS kernels $K^{(\rm inst)}_{\rm CS}$ to allow for evolution at higher resolution and rapidities.  The evolved TMDs at higher resolutions and rapidities, show substantial shift to lower parton-$x$.  Interpolating the pQCD results at small $b_\perp$ and  the ILM results at large $b_\perp$ for the rho TMDs, allows the extraction of the TMDs in transverse space for fixed parton-$x$. 

The rho meson TMDs integrated over $k_\perp$ yield its four PDFs. The resulting rho tensor is about 1
at low resolution, which is consistent with the constituent quark model, as well as lattice~\cite{Zhang:2024plq} and model estimate~\cite{Ninomiya:2017ggn}. The quark spin sum  or axial charge is also about 1 at low resolution, in contrast to lattice~\cite{Zhang:2024plq} and model estimates~\cite{Ninomiya:2017ggn} which puts it near $\frac 12$.  This is notable as it implies 
consistency with the constituent quark model for a rho meson in the ILM. 
This may not be surprising,
given the $LL+RR$ character of this meson, which does not mix with the $LR$ fluctuations strongly induced by the instantons and anti-instantons (but mix with their topologically trivial molecules),
which are a major source of intrinsic spin in the pion and nucleon. This point  deserves further
investigations, namely on a lattice with different resolutions.

Indeed, the emergent tomographic picture of the
rho meson at different resolutions is encouraging for future comparison with lattice simulations, 
assuming that the rho meson state can be numerically separated from the mixing with the low-lying multi-pion states. Using the gradient flow technique, a direct comparison to our TMDs at low resolution would be very useful.

Finally, the extension of the present analysis to include the 4-Fock space
contribution with possible P-wave admixing, may prove important in understanding the possible role of the orbital contribution in the rho meson TMDs.

\vskip 1cm
{\bf Acknowledgments\,\,}
This work  is supported by the Office of Science, U.S. Department of Energy under Contract  No. DE-FG88ER40388. This research is also supported in part within the framework of the Quark-Gluon Tomography (QGT) Topical Collaboration, under contract no. DE-SC0023646.

\appendix

\section{Polarization and spin $4$-vectors}
For a spin-$j$ particle with a mass $m$ in the rest frame, the spin operator $\vec{\Sigma}$ is defined by a $(2j+1)\times(2j+1)$ matrix in the irreducible representation 
of the spin algebra, with the unit vector $\vec{s}$ as the (arbitrary) spin quantization axis. In a boosted frame,  the rest frame $4$-vector $(0,\vec{s})$ transforms to
\begin{equation}
\label{spin-vec}
    S^\mu=\left(\frac{\vec{p}\cdot\vec s}{m}, \vec{s}+\frac{(\vec{p}\cdot\vec s) \vec p}{m(E_{\vec p}+m)}\right)
\end{equation}
with the condition $p\cdot S=0$ and $S^2=-1$ alongside. 
The spin operator $\vec\Sigma$ is then the Pauli-Lubanski pseudo-vector $\Sigma^\mu$.
\begin{equation}
    \Sigma^\mu=\frac1{2m}\epsilon^{\mu\nu\rho\sigma}p_\nu J_{\rho\sigma}
\end{equation}
where $J_{\mu\nu}$ are the generators of the Lorentz group.


For a spin-$1$ particle with a mass $m_\rho$, the spin operator is a $3\times 3$ matrix with entries $(\Sigma^k)_{ij}=-i\epsilon_{kij}$, defining the spin-$1$ eigenstates $\vec\varepsilon_\lambda$ by
\begin{equation}
\label{eq:spin-1}
-is^k\epsilon_{kij}~\varepsilon^j_\lambda=\lambda~\varepsilon^i_\lambda
\end{equation}
with the polarization $\lambda=0, \pm1$. If we set the spin quantization axis in the  $z$-direction, the polarization vectors are $\vec\varepsilon_\pm=(1,\pm i,0)/\sqrt{2}$ and $\vec\varepsilon_0=(0,0,1)$.

The covariant generalization of  \eqref{eq:spin-1} is 
\begin{equation}
    \frac{i}{m_\rho}\epsilon_{\mu\nu\rho\sigma}S^\mu p^\nu \epsilon^\sigma_\lambda(p)=\lambda\epsilon^\rho_\lambda(p)
\end{equation}
with a $4$-vector generalization of the spin-$1$ eigenstate $\vec\varepsilon_\lambda$. The polarization $4$-vectors of a spin-one particle are given by
\begin{equation}
\varepsilon^\mu_\lambda(p)=\left(\frac{\vec{p}\cdot\vec{\varepsilon_\lambda}}{{m_\rho}},\vec{\varepsilon}_\lambda+\frac{\vec{p}(\vec{p}\cdot\vec{\varepsilon}_\lambda)}{m_\rho(E_{\vec{p}} +m_\rho)}\right)
\end{equation}
with the Ward identity $p\cdot\varepsilon_\lambda(p)=0$ and normalization condition $\varepsilon^*_{\pm}\cdot\varepsilon_\pm=\varepsilon^*_{0}\cdot\varepsilon_0=-1$. The following relation between
the polarization and spin 4-vectors \eqref{spin-vec} is then 
\begin{equation}
\label{spin_exp}
\frac{i}{m_\rho}\epsilon^{\mu\nu\rho\sigma}
p_\nu\varepsilon^*_{\lambda\rho}(p)\varepsilon_{\lambda\sigma}(p)=\lambda S^\mu
\end{equation}

It is convenient to recast the covariant form of the spin-$1$ density matrix $\varepsilon^i_\lambda\varepsilon^j_\lambda{}^*$ for pure spin states with some
fixed polarization $\lambda$, using the spin and momentum 4-vectors. The covariant spin density matrix with a fixed $\lambda$ pure state can be expressed as
\begin{widetext}
\begin{equation}
\varepsilon^{\mu*}_\lambda(p)\varepsilon^{\nu}_\lambda(p)=\frac13\left(-g^{\mu\nu}+\frac{p^\mu p^\nu}{m^2_\rho}\right)-\frac{i\lambda}{2m_\rho}\epsilon^{\mu\nu\rho\sigma}p_\rho S_\sigma-\frac{3\lambda^2-2}{2}\left[S^\mu S^\nu-\frac13\left(-g^{\mu\nu}+\frac{p^\mu p^\nu}{m^2_\rho}\right)\right]
\end{equation}
\end{widetext}
The identity is achieved by decomposing the spin-$1$ density matrix into symmetric traceless and antisymmetric Lorentz tensors and trace. 
using \eqref{spin_exp} and the
completeness relation of polarization vector $\varepsilon^\mu_\lambda(p)$ \cite{Jaffe1997,Jaffe:1988up}.


For simplicity, here we choose the 3-momentum of the target along the $z$ direction ($p_\perp=0$) and denote the Cartesian components of the spin quantization
unit vector by $\vec s= (s_\perp,s_L)$, where $s_\perp = (s^1,s^2)$ is normal to the hadron target momentum and $s_L$ is the $z$ component. In the light-cone signature, the spin vector can be written as
\begin{equation}
\label{spin-vec}
    S^\mu=s_L\frac{p^+}{m_\rho}\bar n^\mu-s_L\frac {m_\rho}{2p^+} n^\mu+s^\mu_\perp
\end{equation}
where $s_L$ can be viewed as the target averaged helicity and the transverse spin is defined by $s_\perp^2=1-s_L^2$.

Generally, for any given direction $\vec S$, the spin of the particle can be projected onto this direction. We note that longitudinal polarization means that $s_\perp$ = 0 and $|s_L|=1$. There  are $2j+1$ spin projections for a spin-$j$ particle along the direction of the momentum. Transverse polarization means that $s_L = 0$ and $|s_\perp| = 1$, and that there
are $2j+1$ spin projections along the direction $s_\perp$ perpendicular to the momentum.

\section{Spin-dependent Wave Functions on the Light Front}
\label{Appx:LFspinor}
Throughout this paper, our conventions of the light front frame follows Kogut-Soper convention based on the Weyl chiral basis of the gamma matrices
\begin{eqnarray}
&\gamma^0=\begin{pmatrix}
0 & \mathds{1} \\
\mathds{1} & 0 \\
\end{pmatrix} ~\
&\gamma^{i}=\begin{pmatrix}
0 & \sigma^i \\
-\sigma^i  & 0 \\
\end{pmatrix}
\end{eqnarray}
The light front components are normalized to
\begin{equation}
\gamma^\pm=\frac{\gamma^0\pm \gamma^3}{\sqrt{2}}
\end{equation}
with the light front projectors defined as
\begin{equation}
\mathcal{P}_+=\frac{1}{2}\gamma^-\gamma^+=\begin{pmatrix}
1 & 0 & 0 & 0 \\
0 & 0 & 0 & 0 \\
0 & 0 & 0 & 0 \\
0 & 0 & 0 & 1 \\
\end{pmatrix}
\end{equation}
and $\mathcal{P}_-=1-\mathcal{P}_+$.

The spin-dependent wave functions denotes the spin states in the creation of a quark-anti-quark pair.  The spin wave functions for each channels can be computed by the light front spinors for the quarks and anti-quarks. The spinors are defined as 
\begin{equation}
u_s(k)=\frac{1}{\sqrt{\sqrt{2}k^+}}\left(\gamma^-k^+-\gamma_\perp\cdot k_\perp+M\right)\begin{pmatrix}\xi_s \\ \xi_s \end{pmatrix}
\end{equation}
and
\begin{equation}
v_s(k)=\frac{1}{\sqrt{\sqrt{2}k^+}}\left(\gamma^-k^+-\gamma_\perp\cdot k_\perp-M\right)\begin{pmatrix}-\eta_s \\ \eta_s \end{pmatrix}
\end{equation}
respectively, 
with $\xi_s$ ($\eta_s$), a quark 2-spinor (anti-quark 2-spinor) with a spin pointing in the $z$-direction, and $M$ the constituent quark mass. The anti-quark 2-spinor is related to a quark 2-spinor by charge conjugation $\eta_s=i\sigma_2\xi_s^*=-2s\xi_{-s}$. The corresponding $\rho$ meson spin wave functions with a meson momentum $p$ along $z$ direction read

\begin{equation}
\begin{aligned}
   &\bar{u}_{s_1}(k_1)\frac{\gamma^1+i\gamma^2}{\sqrt{2}} v_{s_2}(k_2)\\
   =&
        \frac{-1}{\sqrt{x\bar{x}}}\xi_{s_1}^\dagger\left[\sqrt{2}M\sigma^++\left(\bar{x}\frac{1+\sigma_z}{\sqrt{2}}+x\frac{1-\sigma_z}{\sqrt{2}}\right)k_R\right]\eta_{s_2}
\end{aligned}
\end{equation}

\begin{equation}
\begin{aligned}
   &\bar{u}_{s_1}(k_1)\frac{\gamma^1-i\gamma^2}{\sqrt{2}} v_{s_2}(k_2)\\
   =&
        \frac{-1}{\sqrt{x\bar{x}}}\xi_{s_1}^\dagger\left[\sqrt{2}M\sigma^--\left(\bar{x}\frac{1-\sigma_z}{\sqrt{2}}+x\frac{1+\sigma_z}{\sqrt{2}}\right)k_L\right]\eta_{s_2}
\end{aligned}
\end{equation}

\begin{equation}
\begin{aligned}
   \bar{u}_{s_1}(k_1)\gamma^+ v_{s_2}(k_2)=2\sqrt{x\bar{x}}\, p^+\xi_{s_1}^\dagger\sigma_z\eta_{s_2}
\end{aligned}
\end{equation}

\begin{equation}
\begin{aligned}
   \bar{u}_{s_1}(k_1)\gamma^- v_{s_2}(k_2)=-\sqrt{x\bar{x}}\,\left(\frac{k^2_\perp+M^2}{x\bar{x}p^+}\right)\xi_{s_1}^\dagger\sigma_z\eta_{s_2}
\end{aligned}
\end{equation}
where $\sigma^\pm=(\sigma^1\pm i\sigma^2)/2$ and $k_{L,R}=k^1_\perp\pm ik^2_\perp$

\section{Summary of the rho Wavefunctions in the ILM}
\label{rho_LFWF}
The twist-2 distribution amplitude (DA) for the longitudinally polarized  rho meson is
\begin{widetext}
\begin{equation}
\label{Rho_DAl}
\varphi_{\rho,\mathbin{\|}}(x)=
\frac  1{f_\rho}\int_{-\infty} ^{+\infty} \frac{dz^-}{2\pi}e^{ixp^+ z^-}\langle0|\overline{q}(0)\frac{\tau^a}2\gamma^+W[0,z^-]q(z^-)|\rho(\lambda=0)\rangle
\end{equation}
The twist-3 distribution amplitude (DA) for the transversely polarized rho meson is
\begin{eqnarray}
\label{Rho_DAt}
\varphi_{\rho,\perp}(x)\epsilon^i_\perp(p)=
\frac  {p^+}{m_\rho f_\rho }\int_{-\infty} ^{+\infty}  \frac{dz^-}{2\pi}e^{ixp^+ z^-}\langle0|\overline{q}(0)\frac{\tau^a}{2}\gamma^i_\perp W[0,z^-]q(z^-)|\rho(\lambda=\pm1)\rangle\nonumber\\
\end{eqnarray}
Here $f_\rho\approx 155\,\rm MeV$ \cite{ParticleDataGroup:2018ovx} is the 
rho electroweak decay constant. 
The explicit form of the rho DAs in the ILM read~\cite{Liu:2023fpj}
\begin{eqnarray}
\varphi_{\rho,\mathbin{\|}}(x)&&=
\frac{\sqrt{N_c}}{m_\rho f_\rho}\frac{C_{\rho_L}}{8\pi^2}\theta(x\bar{x})\int_0^{\infty} dk^2_\perp  \frac{4(k_\perp^2+M^2)}{k_\perp^2+M^2-x\bar{x}m^2_\rho} \mathcal{F}^2\bigg(\frac{k_\perp}{\lambda_{\rho} \sqrt{x\bar x}}\bigg)\\
\varphi_{\rho,\perp}(x)&&=
\frac{\sqrt{N_c}}{m_\rho f_\rho}\frac{C_{\rho_T}}{8\pi^2}\frac{\theta(x\bar{x})}{x\bar{x}}\int_0^{\infty} dk^2_\perp  \frac{(1-2x\bar x)k_\perp^2+M^2}{k_\perp^2+M^2-x\bar{x}m^2_\rho} \mathcal{F}^2\bigg(\frac{k_\perp}{\lambda_{\rho} \sqrt{x\bar x}}\bigg)
\end{eqnarray}
\end{widetext}
Here $\bar x =1-x$ and the  unit step function $\theta(x\bar{x})=1$ constrains the DAs  to the  physical domain $0<x<1$.  The parameter $\lambda_\rho=1.260$ is  chosen so  that the DAs are properly normalized at $\mu=1/\rho$,
for  a fixed  decay constant $f_\rho$.


\section{DGLAP evolution}
The quark distribution $q(x)$ and gluon distribution $g(x)$ evolves with energy scale, 
\begin{equation}
\begin{aligned}
\frac{dq(x)}{d\ln\mu^2}=&\int^1_x \frac{dx'}{x'}~p_{qq'}\left(\frac x{x'}\right)q(x')+p_{qg}\left(\frac x{x'}\right)g(x')\\
\frac{dg(x)}{d\ln\mu^2}=&\int^1_x \frac{dx'}{x'}~ p_{gq'}\left(\frac x{x'}\right)q(x')+p_{gg}\left(\frac x{x'}\right)g(x')
\end{aligned}
\end{equation}
Partons $q$ or $g$ come out from the hadronic bound state carrying the longitudinal momentum $k^+=x'p^+$ and split into a parton pair carrying momentum $(x/x')k^+$ and $(1-x/x')k^+$ respectively. The evolution kernel can be written in terms of a perturbative expansion
\begin{equation}
    p_{ij}(x)=\sum_n\left(\frac{\alpha_s(\mu)}{4\pi}\right)^{n+1}p^{(n)}_{ij}(x)
\end{equation}
For the parton distribution $q(x)$ and $g(x)$, the splitting functions are defined as
\begin{align}
    p^{(0)}_{qq}(x)=&2C_F\left(\frac{1+x^2}{(1-x)_+}+\frac{3}{2}\delta(1-x)\right)\nonumber\\
    p^{(0)}_{qg}(x)=&N_f\left(x^2+\left(1-x\right)^2\right)\nonumber\\
    p^{(0)}_{gq}(x)=&2C_F\left(\frac{1+(1-x)^2}{x}\right)\nonumber\\
    p^{(0)}_{gg}(x)=&4C_A\left(\frac{x}{(1-x)_+}+\frac{1-x}{x}+x(1-x)\right)\nonumber\\
    &+\beta_0\delta(1-x)
\end{align}
where the one-loop beta function coefficient is $\beta_0=\frac{11}{3}C_A-\frac{2}{3}N_f$, $C_F=\frac{N_c^2-1}{2N_c}$, and $C_A=N_c$$C_A=N_c$

For the helicity distribution $\Delta q(x)$ and $\Delta g(x)$, the evolution is of the same form with similar splitting functions defined by \cite{Moch:2014sna}
\begin{align}
\Delta p^{(0)}_{qq}(x) =& 2C_F\left(\frac{2}{(1-x)_+} - 1 - x+\frac32 \delta(1-x)\right) \nonumber\\
\Delta p^{(0)}_{qg}(x) =& 2N_f(2x - 1)\nonumber\\
\Delta p^{(0)}_{gq}(x) =& 2C_F(2 - x) \nonumber\\
\Delta p^{(0)}_{gg}(x) =& 2C_A\left(\frac{2}{(1-x)_+} + 2 - 4x\right) \nonumber\\
&+\beta_0\delta(1-x)
\end{align}

For the transversity distribution $\delta q(x)$, with  no corresponding gluon distribution, the evolution does not mix with gluon, leaving the evolution kernel with one splitting function defined by \cite{Vogelsang:1997ak,Mikhailov:2008my,Artru:1989zv}
\begin{equation}
    \delta p^{(0)}_{qq}(x) = 2C_F \left( \frac{2}{(1-x)_+} -2+ \frac{3}{2} \delta(1-x) \right)
\end{equation}

\section{Subtracted TMD distributions in $b$-space}
\label{sec:tmd_b}
Here we summarize the Fourier transformation for each of the structure functions in s the ubtracted TMD distributions in \eqref{tmd_b_2}
\begin{align}
\tilde{f}^{q/\rho}_1(x,b)=&2\pi\int_0^\infty dk k f_1^{q/\rho}(x,k)J_0(kb)\nonumber\\
\tilde{f}^{q/\rho}_{1LL}(x,b)=&2\pi\int_0^\infty dk k f_{1LL}^{q/\rho}(x,k)J_0(kb)\nonumber\\
\tilde{f}^{q/\rho}_{1LT}(x,b)=&\frac{2\pi}{m_\rho^2b}\int_0^\infty dk k^2 f_{1LT}^{q/\rho}(x,k)J_1(kb)\nonumber\\
\tilde{f}^{q/\rho}_{1TT}(x,b)=&\frac{2\pi}{m_\rho^4b^2}\int_0^\infty dk k^3 f_{1TT}^{q/\rho}(x,k)J_2(kb)
\end{align}

\begin{align}
\tilde{g}^{q/\rho}_{1}(x,b)=&2\pi\int_0^\infty dk k g_{1}^{q/\rho}(x,k)J_0(kb)\nonumber\\
\tilde{g}^{q/\rho}_{1T}(x,b)=&\frac{2\pi}{m_\rho^2b}\int_0^\infty dk k^2 f_{1LT}^{q/\rho}(x,k)J_1(kb)
\end{align}

\begin{align}
\tilde{h}^{q/\rho}_{1}(x,b)=&2\pi\int_0^\infty dk k h_{1}^{q/\rho}(x,k)J_0(kb)\nonumber\\
\tilde{h}^{q/\rho}_{1L}(x,b)=&\frac{2\pi}{m_\rho^2b}\int_0^\infty dk k^2 h_{1L}^{q/\rho}(x,k)J_1(kb)\nonumber \\
\tilde{h}^{q/\rho}_{1T}(x,b)=&\frac{2\pi}{m_\rho^4b^2}\int_0^\infty dk k^3 h_{1T}^{q/\rho}(x,k)J_2(kb)
\end{align}

\section{CS kernel}
The integrability of the RG evolution equations garantees
\begin{equation}
\label{RG_K}
    \frac{dK_{\rm CS}(b_\perp,\mu) }{d\ln \mu^2}=-\Gamma_{\mathrm{cusp}}(\alpha_s(\mu))
\end{equation}
The solution to the CSS renormalization group equation \eqref{RG_K} reads 
 \begin{equation}
\begin{aligned}
\label{CS_kernel}
    &K_{\rm CS}(b_\perp,\mu)\\
    =&\begin{cases}
        K^{(\rm pert)}_{\rm CS}(b_\perp, \mu_{b})-2\int_{\mu_{b}}^\mu \frac{d\mu'}{\mu'}\Gamma_{\rm cusp}\ ,\ b_\perp\ll b_{\rm NP} \\
        K^{(\rm inst)}_{\rm CS}(b_\perp)\ ,\ b_\perp\gg b_{\rm NP} 
    \end{cases}
\end{aligned}
\end{equation}
The Collins-Soper kernel is split into a perturbative contribution plus  
a non-perturbative remainder. The splitting is enforced  by a smooth cut-off $b_{\rm NP}\sim\rho$ with a transverse scale $$\mu_{b} = \frac{2 e^{-\gamma_E}}{b_\perp}$$ 
to ensure that the evolution remains within the perturbative regime.

The coefficients for each $\alpha_s$ order of the perturbative part of $K^{(\rm pert)}_{\rm CS}$ with minimal logarithmic dependence, can be found in \ref{Appx:cusp}. The perturbative gluon radiation at large transverse distance is suppressed, and the nonperturbative Collins-Soper kernel compensates the gluon radiation non-perturbatively. The $\mu$-independent non-perturbative correction to the Collins-Soper kernel 
can be estimated in the ILM, as we have done recently in~\cite{Liu:2024sqj}.


\section{Anomalous dimensions and Wilson coefficients for TMD evolution}
\label{Appx:cusp}
Here we record the perturbative anomalous dimensions and the Wilson coefficients, for the TMD evolution up to next-to-next-to-next-to-leading order (N$^3$LO)~\cite{Davies:1984hs,Collins:1984kg,Collins:2017oxh}

First, the light-cone cusp anomalous dimension is given by
\begin{equation}
    \Gamma_{\mathrm{cusp}}(\alpha_s)=\sum_{n=1}^\infty \Gamma^{(n)}_{\mathrm{cusp}}(\alpha_s)
\end{equation}
where each order is defined as
\begin{equation}
    \Gamma^{(n)}_{\mathrm{cusp}}(\alpha_s)=\left(\frac{\alpha_s}{4\pi}\right)^n\Gamma_n
\end{equation}
with the $n$th order coefficients  defined as
\begin{equation}
    \Gamma_1=4C_F
\end{equation}

\begin{equation}
\begin{aligned}
    \Gamma_2 =&8 C_F \left[ C_A \left( \frac{67}{18} - \frac{\pi^2}{6} \right) - \frac{5}{9} N_f \right]\\
    =&\frac{1072}{9} - \frac{16 \pi^2}{3} - \frac{160}{27} N_f
\end{aligned}
\end{equation}

\begin{equation}
\begin{aligned}
    \Gamma_3 =& 352 \zeta(3) + \frac{176 \pi^4}{15} - \frac{2144 \pi^2}{9} + 1960 \\
    &+ N_f \left( -\frac{832 \zeta(3)}{9} + \frac{320 \pi^2}{27} - \frac{5104}{27} \right) \\
&- \frac{64}{81} N_f^2
\end{aligned}
\end{equation}

\begin{widetext}
\begin{equation}
\begin{aligned}
    \Gamma_4 =& 
    \Bigg(-1536 \zeta(3)^2 - 704 \pi^2 \zeta(3) + 28032 \zeta(3)\\
    &- \frac{34496 \zeta(5)}{3} 
    + \frac{337112}{9} - \frac{178240 \pi^2}{27} + \frac{3608 \pi^4}{5} - \frac{32528 \pi^6}{945}\Bigg)\\
&+ N_f \Bigg( 
    \frac{1664 \pi^2 \zeta(3)}{9} - \frac{616640 \zeta(3)}{81} + \frac{25472 \zeta(5)}{9} 
    \\
    &- \frac{1377380}{243} + \frac{51680 \pi^2}{81} - \frac{2464 \pi^4}{135} 
\Bigg)\\
&+ N_f^2 \Bigg( 
    \frac{16640 \zeta(3)}{81} + \frac{71500}{729} - \frac{1216 \pi^2}{243} - \frac{416 \pi^4}{405} 
\Bigg)\\
&+ N_f^3 \Bigg( 
    \frac{256 \zeta(3)}{81} - \frac{128}{243} 
\Bigg)\\
\end{aligned}
\end{equation}
\end{widetext}

 The collinear anomalous dimension for the TMD operator is given by
\begin{equation}
    \gamma_{f}(\alpha_s)=\sum_{n=1}^\infty \gamma^{(n)}_{f}(\alpha_s)
\end{equation}
where each order is defined as
\begin{equation}
    \gamma^{(n)}_{f}(\alpha_s)=\left(\frac{\alpha_s}{4\pi}\right)^n\gamma_n
\end{equation}
with the $n$th order coefficients are defined as

\begin{equation}
    \gamma_1 = 6 C_F
\end{equation}

\begin{widetext}
\begin{equation}
    \gamma_2 =C_F^2 \left( 3 - 4 \pi^2 + 48 \zeta(3) \right) + C_F C_A \left( \frac{961}{27} + \frac{11 \pi^2}{3} - 52 \zeta(3) \right) + C_F N_f \left( - \frac{130}{27} - \frac{2 \pi^2}{3} \right)
\end{equation}

\begin{equation}
\begin{aligned}
\gamma_3 =& C_F^2 N_f \left( - \frac{2953}{27} + \frac{26 \pi^2}{9} + \frac{28 \pi^4}{27} - 512 \zeta(3) \right) + C_F N_f^2 \left( - \frac{4834}{729} + \frac{20 \pi^2}{27} + \frac{16 \zeta(3)}{27} \right) \\
&+ C_F^3 \left( 29 + 6 \pi^2 + \frac{16 \pi^4}{5} + 136 \zeta(3) - \frac{32 \pi^2 \zeta(3)}{3} - 480 \zeta(5) \right) \\
&+ C_F^2 C_A \left( \frac{139345}{1458} + \frac{7163 \pi^2}{243} + \frac{83 \pi^4}{27} - \frac{7052 \zeta(3)}{9} + \frac{88 \pi^2 \zeta(3)}{9} + 272 \zeta(5) \right) \\
&+ C_A C_F N_f \left( \frac{17318}{729} - \frac{2594 \pi^2}{243} - \frac{22 \pi^4}{27} + \frac{1928 \zeta(3)}{27} \right) \\
&+ C_A C_F^2 \left( \frac{151}{2} - \frac{410 \pi^2}{9} - \frac{494 \pi^4}{135} + \frac{1688 \zeta(3)}{3} + \frac{16 \pi^2 \zeta(3)}{3} + 240 \zeta(5) \right)
\end{aligned}
\end{equation}
\end{widetext}

The 2D evolution of the transverse momentum dependence are  driven by the anomalous dimensions $\gamma_F$ and $\Gamma_{\rm cusp}$. They are identical for TMD PDFs and TMD fragmentation functions to all orders \cite{Collins:2017oxh}. The anomalous dimension $\gamma_F$ only varies by the quark operator, as  defined in the TMD functions.

The perturbative part of the Collins-Soper kernel at minimal logarithmic scale $K^{(\rm pert)}_{\rm CS}(b, \mu=\mu_b)$ is defined as
\begin{equation}
    K^{(\rm pert)}_{\rm CS}(b, \mu_b)=\sum_{n=1}^\infty\left(\frac{\alpha_s(\mu_b)}{4\pi}\right)^nk_n
\end{equation}

\begin{align}
k_1 =& 0 \\
k_2 =&  8 C_F \left[\left( \frac{7}{2} \zeta(3) - \frac{101}{27} \right) C_A + \frac{14}{27} N_f \right]\nonumber\\
=&-2(-56 \zeta(3) + \frac{1616}{27} - \frac{224}{81} N_f) \\
k_3 =& -2\Bigg( \frac{176 \pi^2 \zeta(3)}{3} - \frac{24656 \zeta(3)}{9} + 1152 \zeta(5) \nonumber\\
&+ \frac{594058}{243} - \frac{6392 \pi^2}{81} - \frac{154 \pi^4}{45} \Bigg)  \nonumber\\
&-2 N_f \left( \frac{7856 \zeta(3)}{81} - \frac{166316}{729} + \frac{824 \pi^2}{243} + \frac{4 \pi^4}{405} \right) \nonumber\\
& -2 N_f^2 \left( \frac{64 \zeta(3)}{27} + \frac{3712}{2187} \right)
\end{align}

With the above in mind, the OPE convolution in the small-$b$ ($b\ll b_{\rm NP}$) TMD evolution \eqref{convo_1} can be written as
\begin{widetext}
\begin{equation}
\begin{aligned}
\label{convo_1}
    \tilde{f}_1^{q/\rho}(x, b; \mu_b,\mu_b^2)&
    \simeq\int_x^1 \frac{dx'}{x'}C^f_{q/q}(\frac{x}{x'},b,\mu_b)f_1^{q'/\rho} (x',\mu_b)\\
    \tilde{g}_1^{q/\rho}(x, b; \mu_b,\mu_b^2)&
    \simeq\int_x^1 \frac{dx'}{x'}C^g_{q/q}(\frac{x}{x'},b,\mu_b)g_1^{q'/\rho} (x',\mu_b)\\
    \tilde{h}_1^{q/\rho}(x, b; \mu_b,\mu_b^2)&
    \simeq\int_x^1 \frac{dx'}{x'}C^h_{q/q}(\frac{x}{x'},b,\mu_b)h_1^{q'/\rho} (x',\mu_b)\\
\end{aligned}
\end{equation}
The Wilson coefficient for each type of TMD distribution is given in \cite{Gutierrez-Reyes:2017glx}

\begin{equation}
C^{f,g}_{q/q}\left(x, b, \mu_b \right)=\delta(1-x)+\frac{\alpha_s}{4\pi}C_F \left[ -\frac{\pi^2}{6} \delta(1 - x) + 2(1 - x) \right]
\end{equation}
and
\begin{equation}
C^{h}_{q/q}\left(x, b, \mu_b \right)=\delta(1-x)+\frac{\alpha_s}{4\pi}C_F \left[ -\frac{\pi^2}{6} \delta(1 - x) \right]
\end{equation}
\end{widetext}

\bibliography{ref,refx}

\end{document}